\documentclass[]{jfm}
\usepackage{bm} \usepackage{amssymb} \usepackage{natbib} \usepackage{stmaryrd}
\usepackage{graphicx} \usepackage{color}
\usepackage{amsmath} 
\usepackage{amsfonts}  
\usepackage{bm}
\usepackage{placeins}
\newcommand{\be}{\begin{equation}}
\newcommand{\ee}{\end{equation}}

\usepackage{soul}
\usepackage{color}

\begin{document}

\title[Natural convection with mixed insulating and conducting  boundary conditions]{Natural convection with mixed insulating and conducting  boundary conditions: low and high Rayleigh numbers regimes}

\author[P. Ripesi, L. Biferale, M. Sbragaglia and A. Wirth]{P. \ns R\ls I\ls P\ls E\ls S\ls I$^1$, L.\ns B\ls I\ls F\ls E\ls R\ls A\ls L\ls E$^{1}$, M.\ns S\ls B\ls R\ls A\ls G\ls A\ls G\ls L\ls I\ls A$^{1}$ and A.\ns W\ls I\ls R\ls T\ls H$^{2}$}

\affiliation{$^{1}$ Department of Physics and INFN, University of Rome Tor Vergata, \\ Via della Ricerca Scientifica 1, 00133 Rome, Italy \\ $^{2}$ CNRS LEGI (UMR 5519) Grenoble, France}

\date{\today} 
\maketitle

\begin{abstract}
We investigate the stability and dynamics of natural convection in two dimensions, subject to inhomogeneous boundary conditions. In particular, we consider a Rayleigh-B\'{e}nard (RB) cell, where the horizontal top boundary  contains a periodic sequence of alternating thermal insulating and conducting patches, and we study the effects of the heterogeneous pattern on the global heat exchange, both at low and high Rayleigh numbers. At low Rayleigh numbers, we determine numerically the transition from a regime characterized by the presence of small convective cells localized at the inhomogeneous boundary to the onset of {\it bulk} convective rolls spanning the entire domain. Such a transition is also controlled analytically in the limit when the boundary pattern length is small compared with the cell vertical size. At higher Rayleigh number, we use numerical simulations based on a lattice Boltzmann method to assess the impact of boundary inhomogeneities on the fully turbulent regime  up to $Ra \sim 10^{10}$. 
\end{abstract}


\section{Introduction}
\begin{figure}
\begin{center}
\includegraphics[scale=1.5]{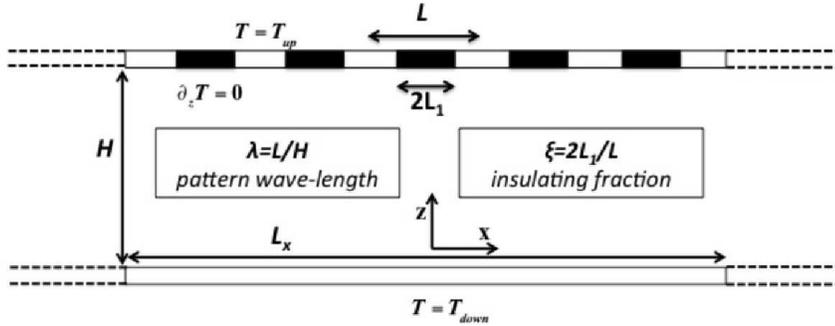}
\vspace{-10mm}
\caption{Sketch of the cell with periodic horizontal boundary conditions and  mixed temperature boundary conditions in the upper wall. Black regions denote insulating properties, white regions denote a constant temperature.}
\label{fig:1}
\end{center}
\end{figure}

Many flows in nature are driven by density differences, they are called convective flows. Thermal convection has applications spanning from cooling devices in micro-computers to heat exchangers in thermal machines. Convection is relevant in biological systems, the earth interior \citep{Guillou,Lenardic1,Lenardic,Jellinek,solomatov}, the ocean \citep{Aargard,Holland,Martinson,wirth}, the atmosphere \citep{Soloviev,cieszelski} and in stars \citep{choudhuri}. In all of these flows many  processes might be involved, as for example rotational effects, phases changes, complex  boundary conditions and non-linear equations of state. Thermal Rayleigh-B\'{e}nard  (RB) convection  is the simplest system of convective motion, see \citet{Bodenschatz, Lohse, Ahlers, Chilla} for recent reviews on the topic. The RB system consists of a fluid subject to an external gravity field with intensity $g$ placed between two horizontal plates,  heated from below and cooled from above.  The associated thermal dynamics is parameterized in terms of  two non-dimensional parameters, namely the Rayleigh number, $Ra=g{\alpha}{\Delta}T {H}^{3}/ (\nu \kappa )$, and the Prandtl number, $Pr=\nu / \kappa $, where $H$ is  the distance between the plates, ${\alpha}$ and $\kappa $ are the thermal expansion and diffusivity coefficients of  the fluid and $\nu$ is the kinematic viscosity. For the standard case when the top and bottom boundaries have homogeneous temperatures the system is known to be linearly unstable and convection starts above a critical Rayleigh number, the latter being determined by the fluid properties and the boundary conditions of the system \citep{chandrasekhar,Rayleigh}. On the other hand, in many, if not all, applications a certain degree of  inhomogeneity is present in the thermal forcing at the boundary. Whenever horizontal inhomogeneities in the boundary conditions appear, as for example a differential heating or cooling within a plate, the system is always unstable as it can be easily checked from the equations of motion. If the inhomogeneities are weak and localized, the convective dynamics is also localized. In the case of  weak and localized inhomogeneities one may still identify a critical Rayleigh number, $Ra_c$, characterizing the transition from the presence of only localized convection in the vicinity of the boundaries to the existence of global {\it bulk} convective motion. The scope of this work is twofold.  First,  we discuss the dependence of such a transition  on the surface heterogeneities.  In particular, we study the onset of large scale thermal convection  in a two dimensional RB cell, the upper plate of which consists  of a periodic sequence of insulating ($\partial_z T=0$) and  ``thermalized'' ($T=T_{up}$) patches, as shown in figure \ref{fig:1}, and quantify the effects of  the boundary heterogeneity on  $Ra_c$. Second, we intend to explore the high Rayleigh numbers regime by changing the pattern length-scale and the Rayleigh number using direct numerical simulations for the same two-dimensional set-up. 
 
The  problem is important for a series of geophysical applications, such as the role of fractures, leads and polynyas in the sea-ice, which is an almost perfect insulator to heat flux. Their existence leads to inhomogeneous convection in both the ocean and the atmosphere. It was recently shown \citep{Marcq} that the size distribution of leads is multi-scale, and that the size matters, small leads (several meters) being more efficient in heat transport than larger ones (several hundreds of meters). The water masses formed by this inhomogeneous convective process around Antarctica are the densest known in the worlds ocean. They sink to the very bottom and are key to the oxygenation of the deepest waters in the ocean and the thermohaline circulation  \citep{Aargard,Holland,Martinson}. The insulating effect of continents on mantle convection in the Earth \citep{Guillou,Lenardic1,Lenardic,Jellinek,solomatov} is another example. Homogeneous RB convection is an important physical problem, but in engineering devices and in nature, inhomogeneities are a conspicuous feature too. Results on inhomogeneous convection is scant which is in stark contrast to the homogeneous case.

In order to make the problem simpler, we specialize here to only one case of inhomogeneities, that is periodically alternating conducting and insulating regions on a one-dimensional pattern at one boundary, the other boundary being homogeneous. Other theoretical, numerical and experimental studies have investigated the onset of convection and the transition to pattern formation in Rayleigh-B\'{e}nard  with periodic temperature modulation on one plate with and without vertical inclination of the cell \citep{Freund,Weiss,Seiden}. More recently,  a detailed study of  transition to bulk convection for a RB cell heated with a sinusoidal profile from below was presented \citep{jfm2013} where some of the issues here discussed are also addressed. In particular, at changing the characteristic wave number of the heating mechanism, the authors study the transition from a system with convection limited to the region close to the boundary condition to a bulk regime with rolls that have different orientation, transversal or longitudinal, depending on the forcing wavenumber.  Our work is distinguishable from the previous ones for at least two reasons. First, we investigate a different set-up, with insulating and conducting regions that cannot be characterized by a single harmonic modulation of the temperature, therefore changing both the type (Dirichlet and Neumann) and spatial characteristics of the boundary conditions with respect to the previous studies. This is clearly inspired by and reflects very well the oceanic context. It also reflects applications to thermal convection at surfaces covered with different materials having different thermal properties as used in many engineering applications. Such boundary condition applies to many applications, but it also involves dynamics of many Fourier modes and their interaction. This makes its analytical and numerical treatment more involved than a sinusoidal (single wave-number) variation of the magnitude of the temperature or the heat-flux. Second, we address also the impact of such modulation on the high Rayleigh number regime addressing the universality of the turbulent statistics at changing the details of the forcing mechanisms.

All numerical simulations have been done using a lattice Boltzmann scheme. Lattice Boltzmann Methods are well known and widely applied to a variety of single and multi-phase hydrodynamic problems \citep{lbm1,lbm2,lbm3,lbm} and they have also been developed to study thermal fluids, both with the Boussinesq approximation  \citep{lbm_ob1,lbm_ob2} and in a fully thermal regime \citep{lbm_NOB1,ZT,lbm_NOB2,SYC06,Philippi06,PrasianakisKarlin07,Gonnellaetal07,Watari09}. Lattice Boltzmann methods are particularly adapted to attack non-homogeneous boundary conditions, thanks to their fully local stream-and-collide nature.  In the following, we first validate the method against exact results in the low Rayleigh number limit and then  we apply it to explore the high Rayleigh number regimes. 

The paper is organized as follows. In section \ref{sec:equations} we discuss the equations describing our problem and we sketch the main idea behind the analytical calculation of $Ra_c$ for the onset of bulk convection. In sections  \ref{s:basicTemp} and \ref{sec:dualSeries} we show analytical and numerical results for the low Rayleigh number conduction of heat. In section \ref{sec:lbm} we briefly summarize a few technical details of the lattice Boltzmann methods, before using it to determine the stability of the system in section \ref{s:stabHggl} and disentangle the high  Rayleigh number regime in section \ref{s:highRa}. Conclusions are given in section \ref{s:concl}.

\section{Non-homogeneous Rayleigh-B\'{e}nard  convection}
\label{sec:equations}
The typical geometrical set-up is depicted in figure \ref{fig:1}, where inhomogeneities are restricted only to the upper plate ($z=H$) and made of alternating regions of either fixed temperature, $T=T_{up}$, or vanishing temperature gradient, $\partial_z T =0$. The lower boundary ($z=0$) is kept at constant temperature, $T=T_{down}$. To study the dynamics of the fluid, we use the classical Oberbeck-Boussinesq \citep{Lohse, Ahlers} equations (repeated indexes are meant summed upon): 
\be
\begin{cases}
\label{eq:NS}
\partial_i u_i= 0\\
\partial_t u_i+u_k \partial_k u_i =-\partial_i P + \nu \partial_{kk} u_i  - \alpha T \, g \delta_{i,z} \\ 
\partial_{t} T+u_i \partial_{i} T  = \kappa \partial_{ii} T,
\end{cases}
\ee
where $u_i$ is the $i$-th component of the velocity field and $P$ the internal pressure of the fluid. We set the origin of coordinates at the lower boundary, and such that the point $(x=0,z=H)$ is located in the middle of one insulating region (see also Figure \ref{fig:1}). The boundary conditions are periodic with a period $L$:
\be
\begin{split}
\begin{cases}
T(x,H)=T_{up} &   x \notin [-L_1+ j L,  L_1+ j L], j \in  \mathbb{Z} \\
\partial_z T(x,H)= 0 & x \in [-L_1+ j L,  L_1+ j L] , j \in \mathbb{Z} \\
T(x,0)=T_{down} & \forall x
\label{eq:bc}
\end{cases}
\end{split}
\ee
where $L_1$ and $L$ are defined in figure \ref{fig:1}. Moreover, we will assume periodic boundary conditions on the horizontal axis and no-slip velocity boundary conditions at both horizontal plates. In this simplified geometry, we have two new control parameters defining the properties of the geometrical pattern, namely the pattern length in units of the cell height, $\lambda = L/H$ and the total percentage of insulating regions, $\xi = 2L_1/L$. In the limiting case $\xi=0$ one recovers the usual RB homogeneous convection, while  $\xi=1$ leads to a purely homogeneous cell with $T=T_{down}$ in the whole domain. Different questions can be asked by changing $\lambda$ and $\xi$. For instance, from an applied point of view,  it is interesting to understand what happens by varying $\lambda$ at fixed $\xi$. This would answer the following question: suppose you have a given percentage of insulating tiles to cover your floor (or ceiling), what is the optimal pattern to reduce/enhance the vertical heat transfer, at fixed temperature jump? Another  important question concerns the horizontal entrainment of turbulent convection inside the stable non-convective regions at different cell heights,  a phenomena that might have important applications to deep convection in the oceans.  From more fundamental aspects, we know that it is difficult to predict the heat flux in the  high Rayleigh number regimes already for the case of purely homogeneous convection. Moreover, the question about universality of large and small scales statistics at changing small details of the forcing and boundary properties is another key issue in turbulence theory and applications. For example, breaking of homogeneity or isotropy in the boundary conditions can affect the flow on a wide range of scales \citep{biferaleprocaccia}. As a result the turbulent statistical properties might be strongly sensitive to symmetry breaking mechanisms and be described by different statistical attractors even in highly turbulent regimes as recently suggested  to explain puzzling transition observed at  high Rayleigh numbers in some experimental set-ups \citep{HeAhlers,Ahlersnjp} and in highly sheared flow \citep{berengere}. Clearly, understanding the effects of possible --small-- boundary heterogeneities for such critical behavior could be key to improve our understanding of such a general question.

\begin{center}
\begin{figure}
\vspace{-100mm}
\hspace{-15mm}
\includegraphics[scale=0.5,angle=-90]{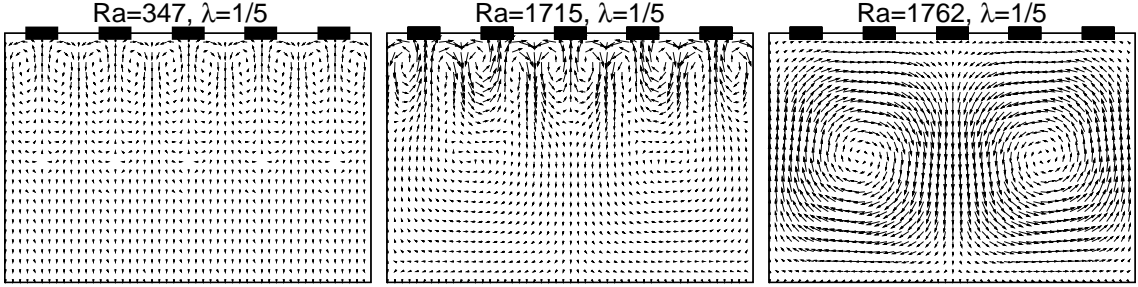}
\caption{Non-homogeneous RB system at $\xi=0.5$, $\lambda=1/5$ at three different Rayleigh numbers, before (left and center panel) and after (right panel) the transition to bulk convection. Vectors for the $Ra=347$ and $Ra=1715$ cases are multiplied by a factor $10$ with respect to the $Ra=1762$ ones. Notice  in the low Rayleigh number regimes, the presence of convective rolls localized at the top boundary before the transition.}
\label{fig:1a}
\end{figure}
\end{center}

\begin{figure}
\begin{center}
\includegraphics[scale=0.26,angle=-90]{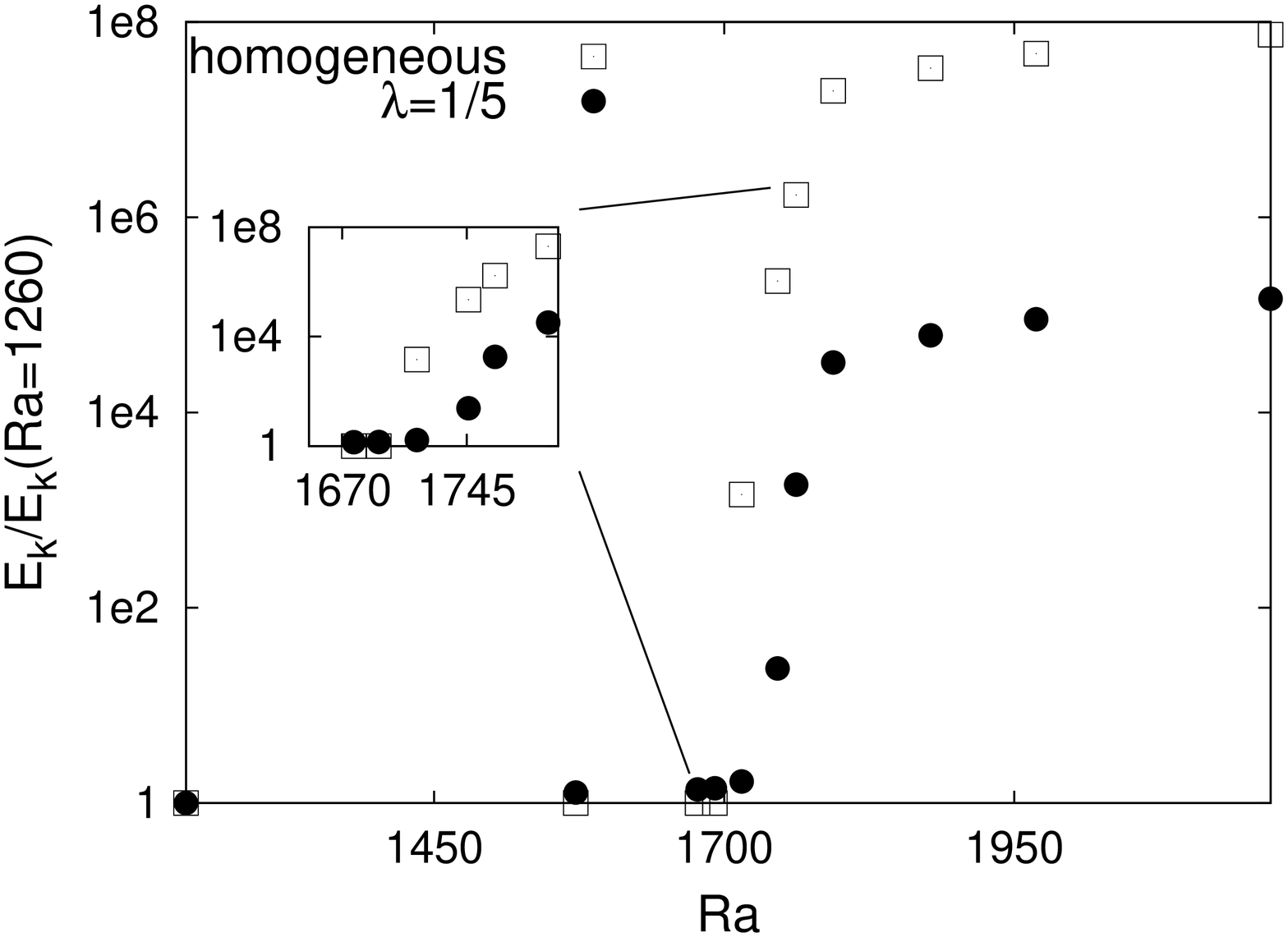}
\includegraphics[scale=0.26,angle=-90]{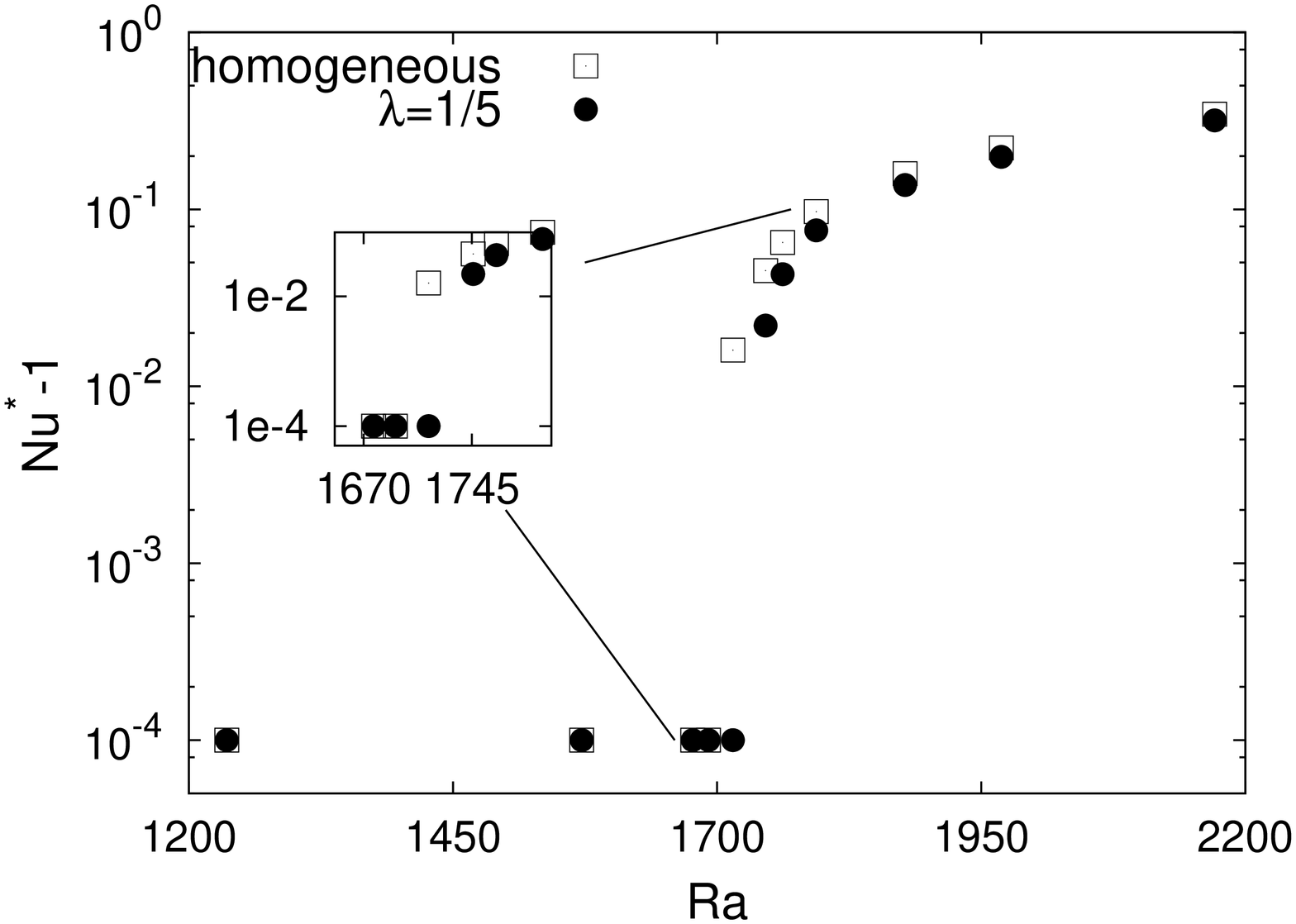}
\caption{Plot of total kinetic energy ($E_k$), compensated with its value at $Ra = 1260$, and Nusselt number ($Nu^*$) as a function of $Ra$ close to the transition for two cases, one homogeneous ($\boxdot$) and one at  $\xi=0.5$ with $\lambda=1/5$ ($\bullet$). Notice the delay in the onset of a global macroscopic convective roll. Inset: enlargement of the region close to $Ra_c$.}\label{fig:transition}
\end{center}
\end{figure}

In the following, we first start to study the low Rayleigh number regime. We know that there exists a critical Rayleigh number for the onset of convection in the homogeneous case. For no-slip velocity boundary conditions it is about $Ra_c(\xi=0) = 1707$ (\citet{chandrasekhar}). In presence of horizontal heterogeneities, the system cannot have a stable static solution with ${\bm u}(x,z,t)=0$ for non vanishing  Rayleigh numbers. So formally $Ra_c(\xi>0) =0$. In fact, the situation is more complex and one may easily imagine that for $\xi \sim 0$ and $\lambda \sim 0$, the situation is not too different from the one of a homogeneous RB cell and that therefore all velocity instabilities are localized close to the top boundary in form of micro-convective cells (see figure \ref{fig:1a}). In other words, we still expect the existence of a critical Rayleigh number characterizing the switch from a near plate micro-convective pattern to a global {\it bulk} convective behavior. Such a transition must be identified in a change of the behavior of the Nusselt number, i.e the  normalized heat flux,  versus the Rayleigh number: 
\be
\label{NUSSELT1}
Nu =\frac{\langle u_z T\rangle_{x,t} -\kappa \partial_z\langle  T\rangle_{x,t} }{\kappa \Delta T/H},
\ee
 where $\Delta T = T_{down} -T_{up}$, as well as in the global behavior of, e.g., the total kinetic energy:
\be
\label{KINETIC}
\qquad E_k  = \frac{1}{2}  \int_0^H \langle u_z^2 + u_x^2 \rangle_x   dz.
\ee 
In the above definitions, we have used $\langle...\rangle_{x,t}$ for the average time and in the $x$ direction, while $\langle...\rangle_{x}$ indicates only the average in the $x$ direction. Let us also notice that one could have adopted a slightly different definition of Nusselt number, taking into account that the solution at zero Rayleigh number, $g=0$,  is not anymore characterized by a homogeneous profile. In this case,  the normalization factor in the denominator must be changed considering that the mean temperature at the upper plate is not anymore given by $T_{up}$. If we denote with $\Delta T_0 (\xi)= 
\langle T(x,H) \rangle_x-T_{down}$ the difference between the mean temperature at the upper plate and the temperature at the bottom plate at zero Rayleigh number, we have: 
\be
\label{NUSSELT2}
Nu^* =\frac{\langle u_z T\rangle_{x,t} -\kappa \partial_z\langle  T\rangle_{x,t} }{\kappa \Delta T_{0}(\xi)/H}.
\ee 
The definition (\ref{NUSSELT2}) ensures that for small enough Rayleigh the Nusselt number tends to $Nu^*=1$  for any $\xi$. In figure \ref{fig:transition} we show the stationary ($t \rightarrow \infty$) value of the total kinetic energy $E_k$ and of the Nusselt number at changing Rayleigh number for two different cases, the classical homogeneous RB ($\xi=0$) and a case with $\xi=0.5$, $\lambda=1/5$. The results have been obtained using a code based on the lattice Boltzmann models (see section 5). As one can notice, the presence of the insulating patches at the top boundary delays the transition to a bulk convection. A rigorous treatment of such phenomena will be discussed in the next section.

\section{Basic Temperature Profile}
\label{s:basicTemp}

In this section we sketch the construction of the basic static temperature profile $T_0(x,z)$ at $g=0$, that is the profile that will define the background configuration also at $g>0$  but for $Ra < Ra_c(\xi)$, i.e. the large scale temperature distribution superposed to the micro-convective cells close to the top boundary. In the bulk region we need to solve a harmonic problem with the boundary conditions given by (\ref{eq:bc}): 
\be
\label{HARMO}
\partial_{xx}  T(x,z)+ \partial_{zz} T(x,z) = 0. \\
\ee
We look for a solution in the form
$$
T(x,z)=T_0(z)+\frac{\Delta T L}{2 \pi H^2}\tilde{\Theta}(x,z),
$$
where 
$$
T_0(z)=T_{down}-\left(\frac{\Delta T}{H}\right) z=T_{down}-\beta z
$$ 
is the usual linear profile with the property that $T_0(0)=T_{down}$ and $T_0(H)=T_{up}$. The boundary condition for the deviations from the linear conductive profile, $\tilde{\Theta}$, are
\be
\begin{split}
\begin{cases}
\tilde{\Theta}(x,H)=0 & L_1<x<\frac{L}{2} \nonumber \\
\partial_z \tilde{\Theta}(x,H)= \frac{2 \pi H}{L} & 0< x < L_1 \nonumber \\
\tilde{\Theta}(x,0)=0 .
\end{cases}
\end{split}
\ee
Because of the symmetry of the problem, we consider only solutions periodic and even for $ x \rightarrow -x$ and we will solve only in the half semi-plane $x\ge 0$. Since $T(x,z)$ is harmonic and $T_0(z)$ is linear, $\tilde{\Theta}(x,z)$ must also be harmonic: $\partial_{xx}\tilde{\Theta}+\partial_{zz}\tilde{\Theta}=0$. Due to the periodicity and the symmetry of the problem, we seek the solution in the following form
$$
\tilde{\Theta}(x,z)=F_0(z)+\sum_{j=1}^{\infty} F_j(z)  \cos \left( \frac{2 \pi j}{L} x \right).
$$
The requirement that eq. (\ref{HARMO}) and the boundary conditions at the lower boundary are satisfied leads to 
$$
\tilde{\Theta}(x,z)=b_0 z+\sum_{j=1}^{\infty} b_j \left(e^{+\frac{2 \pi j z}{L}}- e^{-\frac{2 \pi j z}{L}} \right)  \cos \left( \frac{2 \pi j}{L} x \right)
$$
where $b_0$ and $b_j$ are constants to be fixed upon imposition of the upper plate boundary conditions. We then define the variables $\tilde{x}=\frac{2 \pi}{L} x$  to get:
\be\label{DUALEQS}
\begin{cases}
\tilde{\Theta}(\tilde x,z)=b_0 z+\sum_{j=1}^{\infty} b_j \left(e^{+\frac{2 \pi j z}{L}}- e^{-\frac{2 \pi j z}{L}} \right)  \cos \left( j\tilde x \right)\\
\partial_z \tilde{\Theta}(x,z) =b_0 + \sum_{j=1}^{\infty} b_j \frac{2 \pi j}{L} \left(e^{+\frac{2 \pi j z}{L}}+ e^{-\frac{2 \pi j z}{L}} \right)  \cos \left(j \tilde x \right).
\end{cases}
\ee
It is then possible to verify that  the mixed boundary condition at the upper boundary is given by: 
\be\label{DUALSERIES}
\begin{split}
\begin{cases}
\frac{a_0}{2} +\sum_{j=1}^{\infty} a_j \left(1- e^{-\frac{4 \pi j H}{L}} \right)  \cos \left( j \tilde{x} \right)=0 & c<\tilde{x}<\pi \\
a_0 \frac{L}{4 \pi H} + \sum_{j=1}^{\infty} j a_j  \left(1+ e^{-\frac{4 \pi j H}{L}} \right)  \cos \left( j \tilde{x} \right)=1 & 0<\tilde{x}<c
\end{cases}
\end{split}
\ee
where we have set $c=2 \pi \frac{L_1}{L} = \pi \xi$ and used the definition $a_j=\frac{b_j}{H} e^{\frac{2 \pi j H}{L}}$, $a_0=2 b_0$. As we see from eq. (\ref{DUALSERIES}), the imposition of the mixed boundary condition leads to a typical case of Dual Series (DS) (\citet{sneddon66}), which we can handle numerically quite efficiently \citep{StoneLauga}. Moreover, the solution can be analytically found in some asymptotic cases, as discussed in the following subsections. Once we know the $\{a_j\}$ from (\ref{DUALSERIES}), the temperature profile on the whole domain can be written as: 
\be\label{FIELDFINAL}
T(x,z)=T_0(z)+\frac{\Delta T L}{2 \pi H} \left[ \frac{a_0}{2} \frac{z}{H} +\sum_{j=1}^{\infty}  a_j \left(1- e^{-4 \pi j \frac{z}{L} } \right) e^{+\frac{2 \pi j (z-H)}{L}} \cos \left( \frac{2 \pi j}{L} x \right) \right].
\ee
Let us notice that the above expression implies that the averaged profile along the $x$-direction depends only on the $a_0$ coefficient, i.e.
 it is always linear for the conductive case, at $g=0$:
\be
\label{AVERAGE}
\langle T \rangle_x =T_0(z)+\frac{\Delta T L}{2 \pi H^2} \frac{a_0}{2} z.
\ee

\subsection{Small wavenumber limit of the pattern perturbation, $ L \ll H$, $\lambda \rightarrow  0$}
\label{sub1}

If we assume that $L \ll H$,  we see that the modulation terms along $x$ are active only  for distances of the order of  $L$ from the upper wall (at least when $\xi \ll 1$); for distances larger than $L$ every mode is exponentially damped and the profile reduces to
\be
T(z) \approx T_0(z)+\frac{\Delta T L}{2 \pi H^2} \frac{a_0}{2} z,
\ee
meaning that at distances of the order (and larger) of the periodicity length $L$ from the upper wall we tend to perceive only the average value of the temperature corrections.   In the limit $L \ll H$ (with $a_0 L/H \ll 1$) the non-homogeneous term in the second equation of (\ref{DUALSERIES}) can be neglected and the whole expansion  reduces to a particular case of the general expression:
\be\label{DUAL}
\begin{split}
\begin{cases}
\frac{a_0}{2} +\sum_{j=1}^{\infty} a_j  \cos \left( j \tilde{x} \right)=0 & c<\tilde{x}<\pi \\
\sum_{j=1}^{\infty} j a_j   \cos \left( j \tilde{x} \right)=f(\tilde{x}) & 0<\tilde{x}<c
\end{cases}
\end{split}
\ee
with $f(\tilde{x}) =1$. The solution of the general case is (see Appendix A for all the details):
\be
\label{eq:a0_1}
a_0= 4 \mbox{log}\left( \frac{1}{\mbox{cos}(\frac{c}{2})} \right) 
\ee
where the singularity of the expansion for the purely insulating case, $\xi =1$, is only apparent because  (\ref{eq:a0_1}) is obtained  assuming $a_0 L/H \ll 1$  and therefore one cannot send $\xi \rightarrow 1$ at fixed $L/H$ in the above calculations. Finally, it is also possible to explicitly calculate the whole temperature profile in contact with the insulating region at $z=H$ (see Appendix A):
\be
\label{profH}
\frac{2 \pi H}{\Delta T L} (T(x,H)-T_{up}) = \frac{1}{2}a_{0}+\sum_{j=1}^{\infty} a_{j} \mbox{cos}(j \tilde{x})=2  \mbox{arccosh}\left(\frac{\mbox{cos}(\frac{\tilde{x}}{2})}{\mbox{cos}(\frac{\xi \pi}{2})} \right) \hspace{.2in} 0<\tilde{x}<c.
\ee
\subsection{Large wavenumber limit of the pattern perturbation, $ L \gg H$}
\label{sub2}
In the limit $L \gg H$ we have slits (i.e. strongly vertically confined situations). The temperature in the central region has not enough space to develop a non trivial profile and it stays enslaved to the value at the lower boundary.  The temperature profile is therefore expected to be: 
\be
T(x,z)=\left \{ \begin{array}{l l} T_{down}+\left(\frac{T_{up}-T_{down}}{H}\right) z                &  c< \tilde{x}<\pi \\
 T_{down}  & 0<  \tilde{x} <c   \\
\end{array} \right.
\ee
so that
\be
a_0=\frac{4 \pi H^2}{\Delta T L}\langle T(x,H) -T_{up}\rangle_x = \frac{4 \pi H \xi}{L}.
\ee
This argument can be checked directly in the DS eq. (\ref{DUALSERIES}) which, in the limit $H \ll L$, reduces to
\be
\label{a0lim2}
\begin{split}
\begin{cases}
\frac{a_0}{2} +\sum_{j=1}^{\infty} a_j \frac{4 \pi j H}{L}  \cos \left( j \tilde{x} \right)=0 & c< \tilde{x}<\pi \\
a_0 \frac{L}{4 \pi H} + \sum_{j=1}^{\infty} 2 j a_j  \cos \left( j \tilde{x} \right)=1 & 0< \tilde{x}<c
\end{cases}
\end{split}
\ee
or, alternatively:
$$
a_0 \frac{L}{8 \pi H} +\sum_{j=1}^{\infty}  j a_j   \cos \left( j \tilde{x} \right)=\frac{1}{2}\theta(c-\tilde{x})\theta(\tilde{x}) \hspace{.2in} 0< \tilde{x}< \pi
$$
where with $\theta(x)$ we denote the Heaviside function. Eq. (\ref{a0lim2}) can be solved by calculating the inner product on the interval $0\le \tilde{x}\le \pi$ with $\cos (i \tilde{x})$:
\be
\label{eq:a0_2}
a_0=\frac{4 \pi H \xi}{L}; \qquad 
a_j= \frac{\sin(j \pi \xi)}{\pi j^2}   \hspace{.2in} j>0 
\ee
where we have used
$$
\int_0^{\pi} \cos (i \tilde{x}) \cos (j \tilde{x}) d \tilde{x} =\left \{  \begin{array}{l l} 0  & i \neq j  \\
                                                \frac{\pi}{2}  & i=j, i>0 \\
\end{array} \right. ;
\hspace{.2in}
\displaystyle\int_{0}^{c} \mbox{cos} (i \tilde{x}) d \tilde{x}=\left \{  \begin{array}{l l} c  & i = 0   \\
                                                     \frac{1}{i} \sin(ic) & i \neq 0 . \\
\end{array} \right. 
$$

\subsection{The Case $1 \approx \frac{L}{H} \gg e^{-4 \pi H/L}$ (Intermediate case)}
\label{sub3}
In this limit we can ignore the exponential functions $e^{-\frac{4 \pi j H}{L}}$ and we end up with the following DS
\be 
\label{INTERMEDIATE}
\begin{split}
\begin{cases}
\frac{a_0}{2} +\sum_{j=1}^{\infty} a_j  \cos \left( j \tilde{x} \right)=0 & c< \tilde{x}<\pi \\
a_0 \frac{L}{4 \pi H} + \sum_{j=1}^{\infty} j a_j   \cos \left( j \tilde{x} \right)=1 & 0< \tilde{x}<c
\end{cases}
\end{split}
\ee
where we can find again an exact solution for the DS (see Appendix B for details):
\be
\label{eq:a0_3}
a_0=\frac{4  \mbox{log}\left( \frac{1}{\mbox{cos}(\frac{c}{2})} \right)}{1+\frac{L}{\pi H} \mbox{log}\left( \frac{1}{\mbox{cos}(\frac{c}{2})} \right)}.
\ee


\section{Numerically assisted  solution of dual series}
\label{sec:dualSeries}
Let us now attack the most general case, without any approximation. We start from the DS eq. (\ref{DUALSERIES}) which can be rewritten in the whole interval $0\le \tilde{x} \le \pi$ as: 
\be\label{DS:general}
\sum_{j=0}a_{j} F_{j}(\tilde{x}) \cos(j\tilde{x})=G(\tilde{x})
\ee
with $G(\tilde{x})=\theta(\tilde{x})\theta(c-\tilde{x})$ and
\be\label{DS:generalb}
F_{j}(\tilde{x})=\left \{  \begin{array}{l l} \frac{1}{2}\theta(\tilde{x}-c)\theta(\pi-\tilde{x}) + \left( \frac{L}{4 \pi H} \right)  \theta(\tilde{x}) \theta(c-\tilde{x})  & j=0  \\  \left(1- e^{-\frac{4 \pi j H}{L}} \right) \theta(\tilde{x}-c)\theta(\pi-\tilde{x})+ j   \left(1+ e^{-\frac{4 \pi j H}{L}} \right)\theta(\tilde{x})\theta(c-\tilde{x})             & j>0. \\
\end{array} \right.
\ee
To solve this equation for $\{a_j\}$ numerically \citep{StoneLauga}, we can truncate the series at the order $N$ and calculate its inner product on the interval $0\le \tilde{x}\le \pi$ with $\cos (i \tilde{x})$. This is particularly simple because $F_{j}(\tilde{x})$ is a piecewise constant function. At the end of a lengthy but straightforward analysis we need to solve a linear system
\be\label{SYSTEM}
A_{i,j} a_{j}=y_{i}
\ee
where $A_{i,j}$ is a $N \times N$ matrix and $y_i$ is a vector whose details are reported in Appendix C. For the truncated series, calculations are found to converge well above a truncation order $N$ of a few tens (see Appendix C). Choosing $N=500$, we therefore safely ensure the recovery of the solution with an error less than a fraction of a percent.

\begin{center}
\begin{figure}
\includegraphics[scale=0.5]{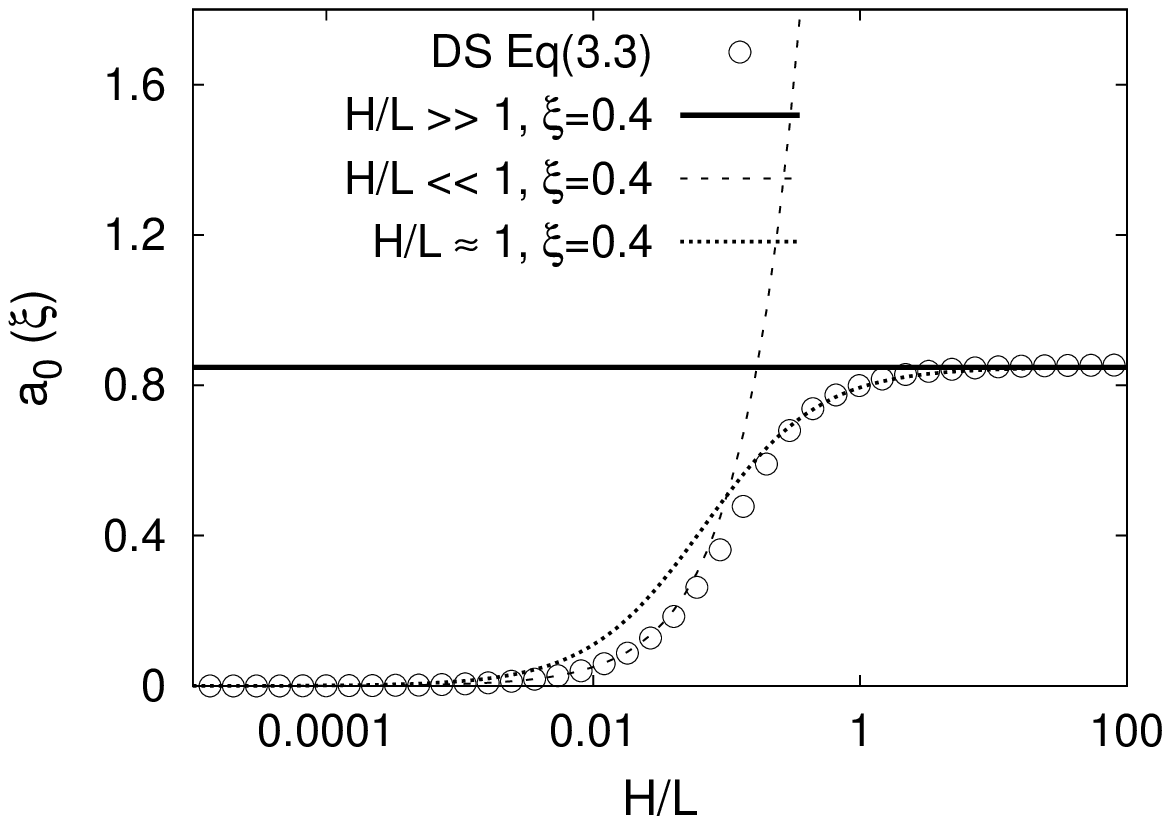}
\includegraphics[scale=0.5]{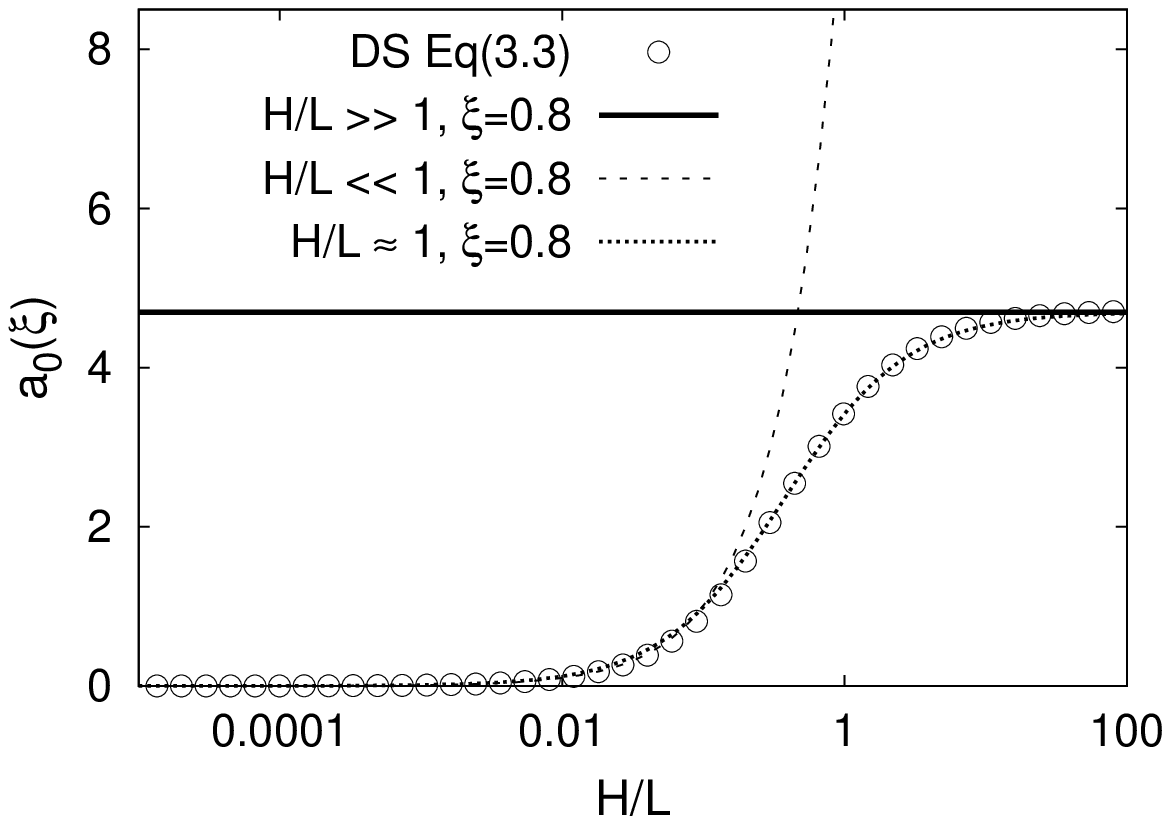}
\caption{Plot of the the zeroth order coefficient, $a_0$, characterizing the mean profile $\langle T \rangle_x$ (see eq. (\ref{AVERAGE})) as extracted from our numerically assisted solution ($\circ$) described in section \ref{sec:dualSeries} for both $\xi=0.4$ (left panel) and $\xi=0.8$ (right panel). The cell height $H$ has been varied from $H \ll L$ to $H \gg L$. The limits $H \gg L$, $H \ll L$ and $1 \approx \frac{L}{H} \gg e^{-4 \pi H/L}$ are reported, as described in eqs. (\ref{eq:a0_1}), (\ref{eq:a0_2}) and (\ref{eq:a0_3}). For the numerical assisted solution of the DS in eq. (\ref{DUALSERIES}), the truncation order is  $N=500$ (see section \ref{sec:dualSeries}).} \label{fig:check1}
\end{figure}
\end{center}

\begin{figure}
\begin{center}
\includegraphics[scale=0.8]{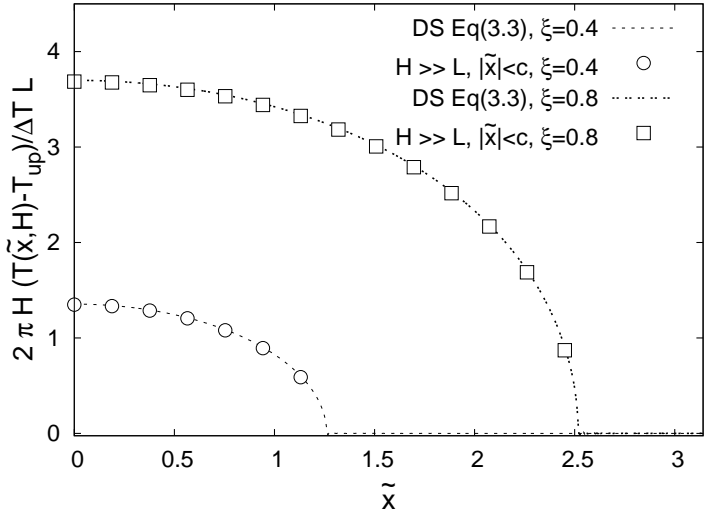}
\caption{Temperature profile in the RB cell with mixed boundary conditions (see figure \ref{fig:1}). Comparison between the analytical solution for the temperature profile at the upper wall ($z=H$) obtained in the limit $H\gg L $, see eq. (\ref{profH}), and the DS solution given by (\ref{DUALSERIES}), numerically evaluated  with $N=500$ as described in section \ref{sec:dualSeries}. The temperature profile is obtained for the  following parameters: $T_{down}=1.5$, $T_{up}=0.5$, $L=2 \pi$ and $H = 100$ for both $\xi=0.4$ ($\circ$) and $\xi=0.8$ ($\boxempty$). The coordinate $x$ is made dimensionless with $2 \pi/L$, i.e.  $\tilde{x}=\frac{2 \pi}{L} x$.}
\label{fig:check2}
\end{center}
\end{figure}

In figure \ref{fig:check1} we plot the coefficient $a_0$ as extracted from our numerically assisted solution of the DS for both $\xi=0.4$ and $\xi=0.8$. $H$ has been varied from $H \ll L$ to $H \gg L$. The expected behavior in the limits discussed in subsections \ref{sub1}-\ref{sub3} $H \gg L$, $H \ll L$ and $1 \approx \frac{L}{H} \gg e^{-4 \pi H/L}$  is also plotted.  As one can see, there is an excellent agreement between the exact numerical solution and the three asymptotic estimates in the limit when they can be applied. Moreover, we notice that the case when we take  $L \sim H $ and neglect the exponential term gives a good first guess for all values of the cell aspect ratio. This is due to the fact that the exponential term is indeed always very small even in the exact solution.  Next, in figure \ref{fig:check2}, we plot the temperature profile at the upper wall as results from the rhs of eq. (\ref{profH}) against the solution of the truncated DS for the  following parameters: $T_{down}=1.5$, $T_{up}=0.5$, $L=2 \pi$ and $H = 100$ for both $\xi=0.4$ and $\xi=0.8$.

\section{Numerical approach based on Lattice Boltzmann Methods}
\label{sec:lbm}

\begin{center}
\begin{figure}
\includegraphics[scale=0.5]{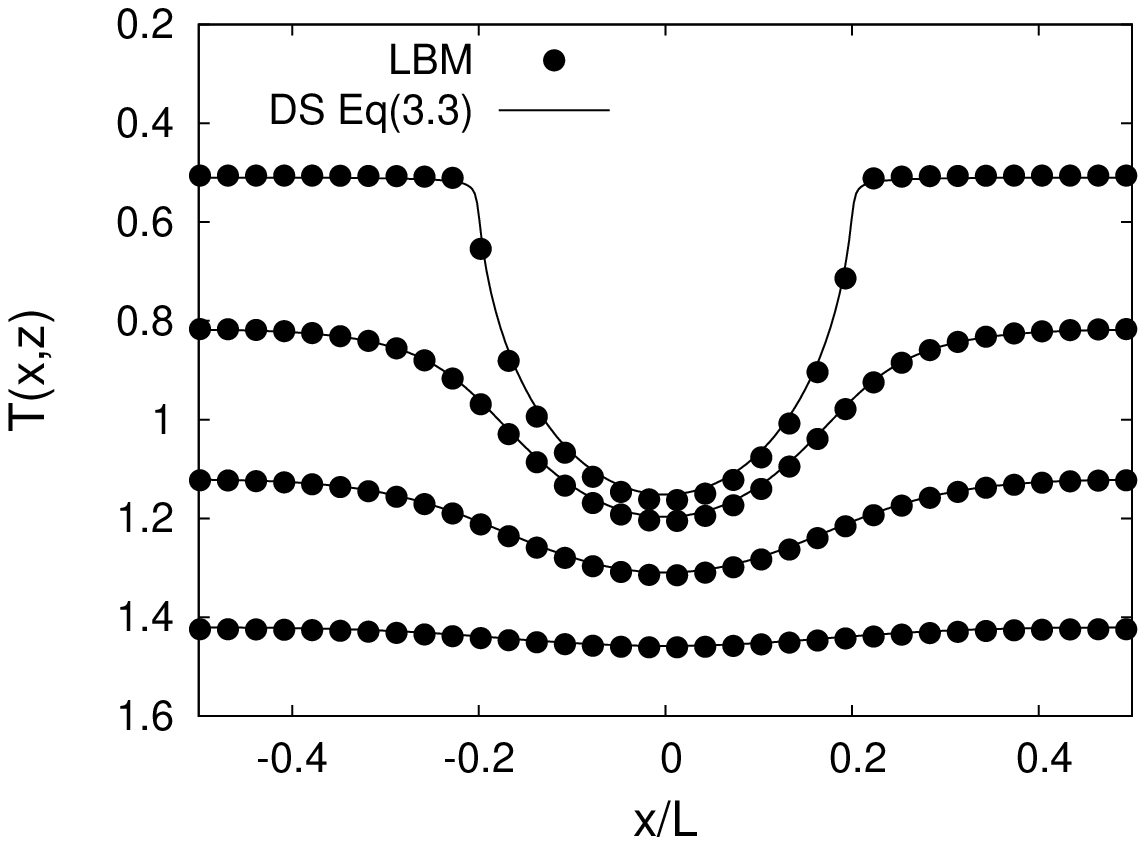}
\includegraphics[scale=0.5]{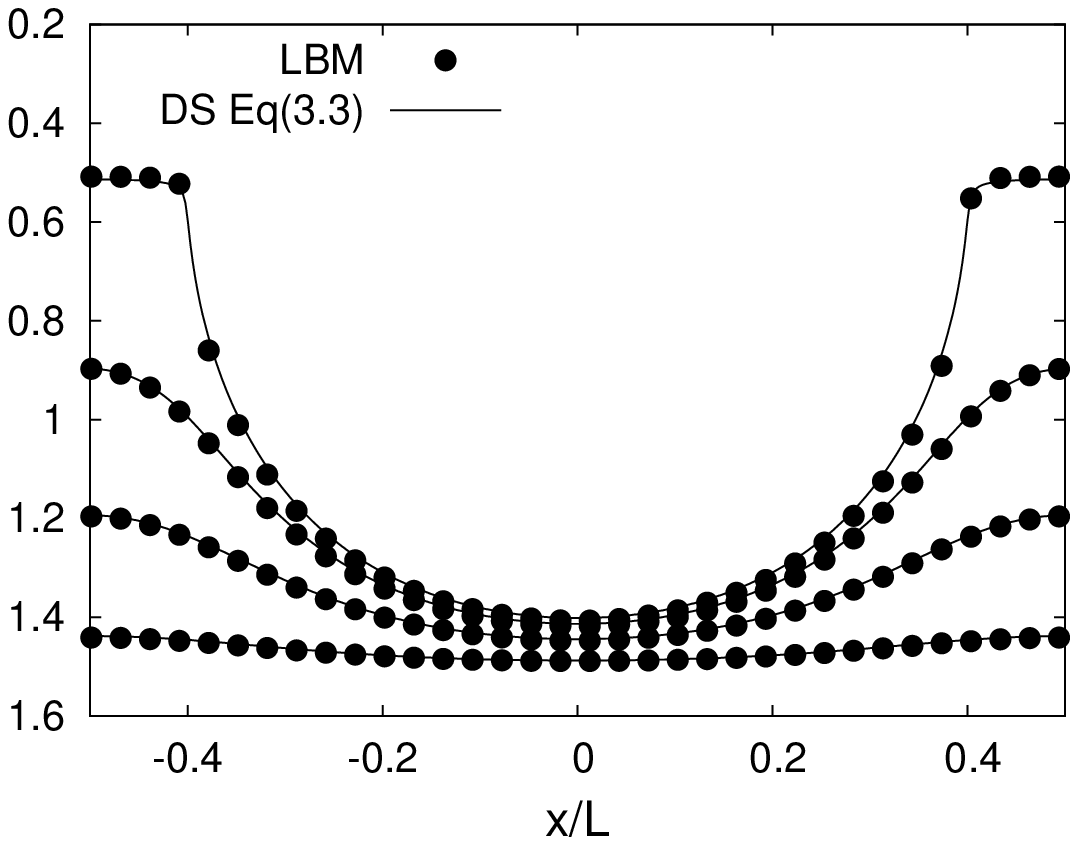}
\caption{Plot of  the temperature field $T(x,z)$ as a function of $x$ for different $z/H=0.1, 0.4, 0.7, 1.0$. We compare lattice Boltzmann simulations ($\bullet$) with the theoretical predictions (solid line) obtained by solving for the $\{a_j\}$ of the DS approach (\ref{DUALSERIES}). The insulating fraction has been chosen to be $\xi=0.4$ (left panel) and $\xi=0.8$ (right panel).\label{fig:check3}}
\end{figure}
\end{center}

\begin{figure}
\begin{center}
\includegraphics[scale=0.8]{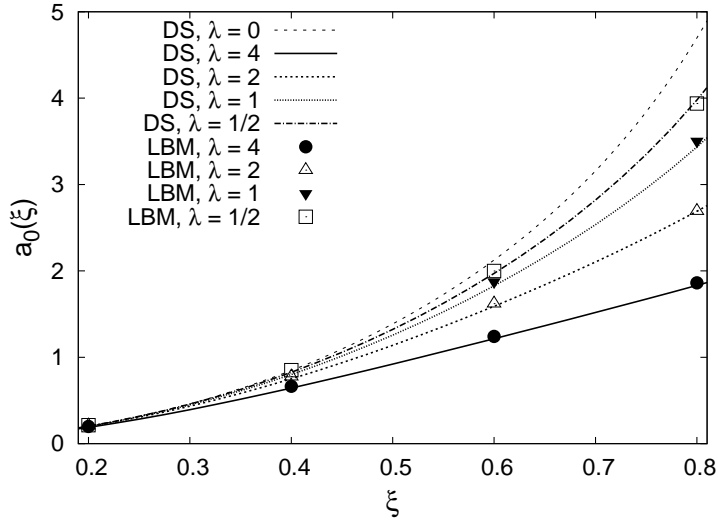}
\caption{The zeroth order coefficient, $a_0$, of the temperature profile expansion  (\ref{FIELDFINAL}). The theoretical predictions have been obtained by solving for the $\{a_j\}$ in the DS approach (\ref{DUALSERIES}) with the numerical procedure illustrated in section \ref{sec:dualSeries}. The lattice Boltzmann prediction is extracted from the average profile consistently with eq. (\ref{AVERAGE}). We have considered the lattice Boltzmann numerical simulations with $L_x=400$, $\tau_f=\tau_g=0.7$ and with the following cell heights: $H=100$ ($\lambda=4$, $\bullet$), $H=200$ ($\lambda=2$, $\vartriangle$), $H=400$ ($\lambda=1$, $\blacktriangledown$), $H=800$ ($\lambda=1/2$, $\boxdot$). The insulating fraction has been varied  between $\xi=0.2$ and $\xi=0.8$. The analytical solution of the DS for $\lambda=0$ (see eq. (\ref{eq:a0_1})) is also reported. \label{fig:check4}}
\end{center}
\end{figure}

In order to go beyond the static cases, we have also developed a numerical algorithm based on the lattice Boltzmann models (LBM) to solve for the whole dynamical problem with the most generic boundary condition. LBM  \citep{Gladrow,BSV,Chen} for ideal isothermal fluids can be derived from the continuum Boltzmann (BGK) equation \citep{BGK54}, upon expansion in Hermite velocity space of the single particle distribution function, $f({\bm x}, {\bm \zeta},t)$, describing the probability of finding a molecule at the space-time location $({\bm x},t)$ and with velocity ${\bm \zeta}$ \citep{HeLuo,ShanHe98,Martys98,Shan06}. Lattice dynamics is enforced with a discrete finite set of velocities ${\bm \zeta} \in [{\bm c}_1, {\bm c}_2,\dots, {\bm c}_M]$, with the total number $M$ determined, case-by-case, by the embedding spatial dimension and the required degree of isotropy  \citep{Gladrow}. As a result, the dynamical  evolution is given by a set of populations $f_{\ell}({\bm x},t)$ with $l=1,\dots, M$ on a discretized spatial and temporal lattice. In what follows we will only address two dimensional cases, where $M=9$ is enough to get the right continuum hydrodynamic limit. As far as we are interested in the Oberbeck-Boussinesq limit, i.e. with thermal properties entering only via a buoyancy term in the Navier-Stokes equations, one may extend the single-fluid LBM dynamics to describe also the evolution of a temperature field by adding another set of populations, $g_{\ell}({\bm x},t)$ (for more complex cases where thermal effects enters also into the equation of state see \citet{lbm_NOB1,lbm_NOB2}). In the two-populations approach the dynamics is then defined by the following  discretized evolution :
\be\label{LBET}
\begin{cases}
f_{\ell}({\bm x}+{\bm c}_{\ell},t+1)-f_{\ell}({\bm x},t)=-\frac{1}{\tau_f}\left(f_{\ell}({\bm x},t)-\bar{f}_{\ell} ({\bm x},t)\right)\\
g_{\ell}({\bm x}+{\bm c}_{\ell},t+1)-g_{\ell}({\bm x},t)=-\frac{1}{\tau_g}\left(g_{\ell}({\bm x},t)-\bar{g}_{\ell}({\bm x},t) \right)
\end{cases}
\ee
where, $\tau_f$ and $\tau_g$ are two characteristic times governing the relaxation dynamics towards the local equilibrium distributions, $\bar{f}_{\ell} ({\bm x},t),\bar{g}_{\ell} ({\bm x},t)$. The hydrodynamic evolution is obtained considering the long wavelength limit (\cite{lbm,Gladrow}) of the equations for the coarse-grained  density, momentum and temperature fields, defined as:
\be\label{DENS}
\rho({\bm x},t)=\sum_{\ell=0}^{8}f_{\ell}({\bm x},t)\qquad \rho {\bm u}({\bm x},t)=\sum_{\ell=0}^{8}{\bm c}_{\ell} f_{\ell}({\bm x},t) \qquad T({\bm x},t)=\sum_{\ell=0}^{8}g_{\ell}({\bm x},t).
\ee
The functional form of the equilibrium for the density-momentum evolution (the first equation of (\ref{LBET})) is given by a discretization of the Maxwellian (repeated indexes are meant summed upon):
\be
\bar{f}_{\ell}({\bm x},t)=\bar{f}_{\ell}(\rho({\bm x},t),{\bm u}^{(S)}({\bm x},t))=w_{\ell}\left[\rho+\frac{\rho u^{(S)}_{k} c^{k}_{\ell}}{c^{2}_{s}}+\frac{(c^{k}_{\ell} c^{s}_{\ell}-c^{2}_{s}\delta_{ks})(\rho u^{(S)}_{k} u^{(S)}_{s})}{2c^{4}_{s}}  \right]
\ee
where $w_{\ell}$ are suitable weights used to enforce isotropy up  to the desired order. The Navier-Stokes equations for the hydro-dynamical velocity field given by the semi-sum of the pre- and post-collision velocity fields,  $ {\bm u}^{(H)}={\bm u}+\frac{\bm F}{2 \rho}$, with the external buoyancy forcing ${\bm F}= \alpha g T {\bm \hat z}$, are then recovered in the Chapman-Enskog limit (with small $g$) if we define the field entering in the local equilibrium by the shifted expression \citep{buickgreated00}: 
\be
 {\bm u}^{(S)} = {\bm u}  + \tau_f \frac{\bm F}{\rho}.
\ee
Concerning the evolution of the temperature field, the local equilibrium (in the second equation of (\ref{LBET})) is given by:
\be
\bar{g}_{\ell}({\bm x},t)=\bar{g}_{\ell}(T({\bm x},t),{\bm u}^{(H)}({\bm x},t))=w_{\ell} T \left[1 + \frac{ u^{(H)}_{k} c^{k}_{\ell}}{c^{2}_{s}}+\frac{(c^{k}_{\ell}c^{s}_{\ell}-c^{2}_{s}\delta_{ks})(u^{(H)}_{k} u^{(H)}_{s})}{2c^{4}_{s}}  \right].
\ee
Let us also notice that in order to get the right hydrodynamic limit of the temperature evolution, an extra body-force term is in principle needed in order to avoid spurious terms in the continuum limit, as shown by \citet{Latt}. The importance of this term depends on the applications. In all our simulations we have checked that it is negligible. In conclusion, in the hydrodynamic limit, one can show that the small Mach number version of the coupled Navier-Stokes equations given by expression (\ref{eq:NS}) is recovered with $\nu = c_s^2(\tau_f -0.5)$ and $\kappa = c_s^2 (\tau_g-0.5)$.  Lattice Boltzmann methods have already been widely used to investigate thermal convection under  different geometries and forcing conditions \citep{lbm_ob1,lbm_ob2},  but never for the case we are focusing here, including non-homogeneous thermal properties at the walls. The locality of the lattice Boltzmann algorithm, allows to enforce the spatial variations in the boundary conditions in a optimal way. To validate the LBM algorithms,  we have run  numerical simulations in a two-dimensional geometry of $L_x \times H$ grid points with   $L_x=400$, $H=100, 200, 400, 800$ and $\tau_f=\tau_g=0.7$ and compared with the analytical results discussed previously.  The insulating  fraction has been varied between $\xi=0.2$ and $\xi=0.8$. The static case ($g=0$, ${\bm u}=0$)  has been reproduced and the temperature profiles are compared with the theoretical prediction in figure \ref{fig:check3}. Similarly, in figure \ref{fig:check4}, we show the comparison between the LBM results and the solution of the DS for the $a_0$ coefficient at changing $\xi$.

\section{The stability of a convective cell in the limit $H \gg L$}
\label{s:stabHggl}

\begin{figure}
\begin{center}
\includegraphics[scale=0.8,angle=0]{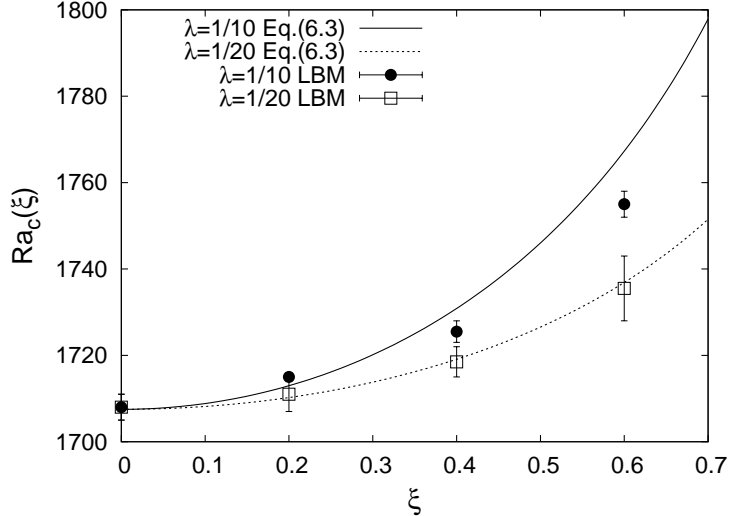}
\caption{Comparison between the prediction for $Ra_c(\xi)$ as given by eq. (\ref{CRITICO}) and the LBM results for $\xi=0.2,0.4,0.6,0.8$ and $\lambda=1/10$ ($\bullet$) and $\lambda=1/20$ ($\boxempty$).  The critical value for $Ra$ is estimated from the transition in the global kinetic energy as shown in figure \ref{fig:transition}. \label{fig:Rac}}
\end{center}
\end{figure}

As we noted in the previous section, when $H \gg L$ (and when $2L_1$ is not close to $L$), the temperature profile (\ref{FIELDFINAL}) simplifies considerably. A modulation along $x$ is present only in a boundary layer of width $\approx L$ close to the upper wall. Away from this boundary layer, the temperature has the form
\be
T(z) \approx T_0(z)+\frac{\Delta T L}{2 \pi H^2} \frac{a_0}{2} z,
\ee
where $\Delta T = (T_{down } -T_{up})$.
Therefore, when we consider in this limit a RB cell with a periodic modulation of normalized length $\lambda$ in the upper boundary condition, we may use the results of an equivalent {\it homogeneous} RB convection but with an effective temperature:
$$
T_{up}^{eff} = T_{up} + \frac{\Delta T L}{2 \pi H} \frac{a_0}{2}. 
$$
Then, we can apply  the same arguments leading to the stability of the RB homogeneous flow provided we redefine the Rayleigh number with a renormalized temperature gradient $\beta^{\prime}=\frac{\Delta T}{H}(1-\frac{a_0 L}{4 \pi H})$
\be
Ra^{\prime}=\frac{g \alpha \beta^{\prime} H^4}{\nu \kappa}=Ra+ Ra  \frac{L}{H \pi} \log \left(\cos \left( \frac{\pi \xi}{2} \right) \right),
\ee
where for $Ra$ we kept the usual definition of the Rayleigh number, i.e. $Ra=\frac{g \alpha \beta H^4}{\nu \kappa}$, of the fully homogeneous set-up. Let us stress again that we are looking here for the critical Rayleigh number at which we should observe a transition from 'localized' convective cells in the belt within a distance $L $ from the upper plate to a {\it bulk} convection.  The criteria for stability in $Ra^{\prime}$ should be unchanged. The convective cell is linearly destabilized when $Ra^{\prime} > Ra^{\prime}_{c} =1707$ (when periodic boundary conditions in the $x$-direction are considered, the horizontal to vertical aspect ratio of the cell is set to two and no-slip boundary conditions for the velocity fields are applied at the two horizontal walls). Translating the result in terms of $Ra$, we identify a critical Rayleigh number given by
\be
\label{CRITICO}
Ra_c(\xi)=\frac{Ra^{\prime}_{c}}{1-\frac{L}{4 \pi H} a_0}=\frac{Ra^{\prime}_{c}}{1+\frac{L}{\pi H}\log \left(\cos \left( \frac{\pi \xi}{2} \right) \right)}.  
\ee
Since $\log \left(\cos \left( \frac{\pi \xi}{2} \right)\right) \le 0$   we see that the flow is stabilized by 
the mixed boundary condition. Let us also note that the  divergence at $\xi=1$ is only apparent, due to the assumption $a_0 L/H \ll 1$ needed to get to (\ref{CRITICO}). Keeping in mind Eq. (\ref{CRITICO}), we have performed numerical simulations for various $\xi$ in order to validate the theoretical argument. Using the  Thermal lattice Boltzmann numerical scheme on a $2D$ domain of size $L_x\times H= 2080 {\times} 1040$, with periodic boundary conditions on lateral walls and no-slip boundary conditions on the upper and lower wall, we have estimated the numerical values of $Ra_c$ at $\xi=$[0, 0.2, 0.4, 0.6] for two configurations, with $\lambda =1/10, 1/20$ respectively. As one can see in figure \ref{fig:Rac}, the agreement between the low $\lambda$-limit (\ref{CRITICO}) and the numerics is good for $\lambda=1/20$, while at smaller separation the effects of the insulating regions enter too much in the bulk and  the prediction (\ref{CRITICO}) is lost. 


In conclusion, we have shown that the mixed boundary condition enhances the average temperature of the top boundary, thus decreasing the average buoyant force in the Navier-Stokes equations. The problem can be paralleled to that of a shear flow over a plate with a regular array of longitudinal no-shear slots 
\citep{philip}, where the effect of the patterning is to provide an effective slip velocity, i.e. an increase of the top temperature in our language. The effective temperature gradient (and therefore the effective Rayleigh number) is decreased as compared to a situation with homogeneous top boundary temperature. Consequently, the critical Rayleigh number at the onset of large scale convection is increased.

\section{Non-homogeneous RB analysis in the high Rayleigh number regime}
\label{s:highRa}

\begin{center}
\begin{figure*}
\includegraphics[scale=.8,angle=0]{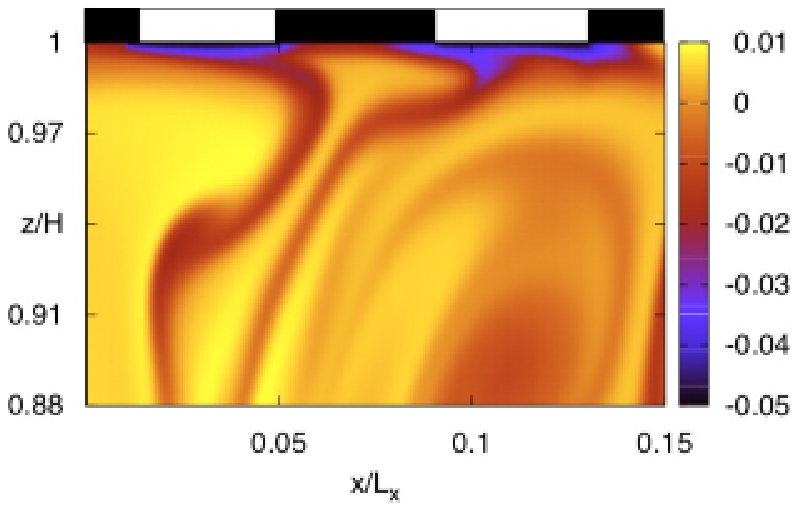}
\includegraphics[scale=.8,angle=0]{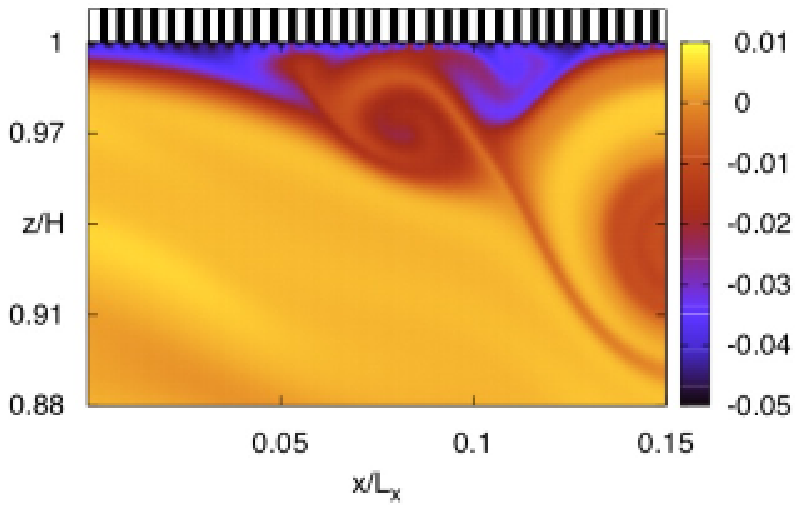}
\caption{(Color online) Two instantaneous snapshots of the temperature field close to the non-homogeneous boundary for $\xi=0.5$ and  two periods of the boundary pattern, $\lambda=1/13$ (left panel), $\lambda=1/208$ (right panel)\label{fig:Tprofile1}}
\end{figure*}
\end{center}

In this section we investigate the high Rayleigh number regime of the RB non homogeneous system. In order to minimize the complexity we will attack only the two-dimensional problem with a fixed percentage of insulating region ($\xi=0.5$) at changing both Rayleigh number and the typical normalized length, $\lambda = L/H$,  of the boundary pattern. A couple of snapshots of the temperature distribution close to the  non-homogeneous boundary  for two cases with large and small $\lambda$  are shown in figure \ref{fig:Tprofile1}. From this qualitative figure one can see  that  when the boundary perturbation is larger than the typical plume size, the thermal activity is concentrated on the conducting regions. On the other hand, when the pattern has a very small length, the presence of cold plumes at the top boundary is fully uncorrelated on the boundary perturbation. Our main goal here is to understand the effects of this boundary 'corrugation' in the conducting properties on the bulk heat transfer at varying Rayleigh number. It is known that other types of corrugation, e.g. induced by a geometrical roughness \citep{castaing}, may lead to non-trivial changes in the global heat transfer. In particular, in the latter case, it is observed that whenever the typical length-scale of the roughness becomes larger than the thermal boundary layer, a transition towards an increased  heat transfer is produced. This can be justified in terms of an effective increase of the plate boundaries which in turns produce a better heat exchange between the boundaries and the bulk fluid. In our set up, an increase of  $\lambda$ above the typical thermal boundary layer length, $\lambda_T$, at a fixed Rayleigh number should in principle lead to the opposite behavior, i.e. to a decrease of the Nusselt number (\ref{NUSSELT1}). This expectation is triggered by the observation that whenever the boundary insulating pattern develops on a length-scale that is larger than  $\lambda_T$  the bulk flow sees a real corrugation in the boundary even in presence of strong turbulent fluctuations. Such corrugation is connected to the presence of regions where the system does not transfer heat, the mean local temperature increases and the global heat exchange should decrease. \\


\begin{figure}
\begin{center}
\includegraphics[scale=0.45,angle=-90]{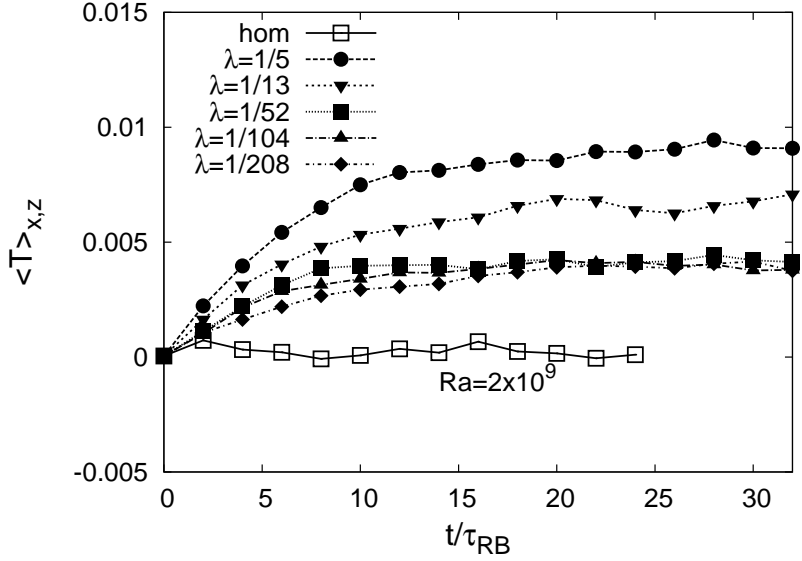}
\caption{Time evolution of the volume averaged temperature for $Ra=2\times{10^9}$ at various $\lambda$ for fixed $\xi=0.5$: $\lambda=1/5$ ($\bullet$), $\lambda=1/13$ ($\blacktriangledown$), $\lambda=1/52$ ($\blacksquare$),  $\lambda=1/104$ ($\blacktriangle$),  $\lambda=1/208$ ($\blacklozenge$), homogeneous ($\boxdot$).}
\label{fig:Tprofile}
\end{center}
\end{figure}

In figure \ref{fig:Tprofile} we plot the time evolution of the volume averaged temperature $\langle T \rangle_{x,z}$ for a given Rayleigh number at changing the pattern periodicity. The initial configuration is given by the unstable homogeneous profile with $T(x,z) = T_{down}$ in the upper half volume and with $T(x,z) = T_{up}$ in the lower half volume, such as the system starts with a Rayleigh-Taylor instability and then tends to develop the (non-homogeneous) RB mean profile. We notice that the mean temperature becomes larger and larger by increasing  $\lambda$, this is clearly due to the fact that for $\lambda > \lambda_{T}$ the insulating regions drives the temperature dynamics in the bulk leading to a net increase in the mean temperature. Because the mean temperature profile must always be symmetric with respect to the center of the cell, it implies that the heat flux decreases.  


\begin{figure}
\begin{center}
\includegraphics[scale=0.8,angle=0]{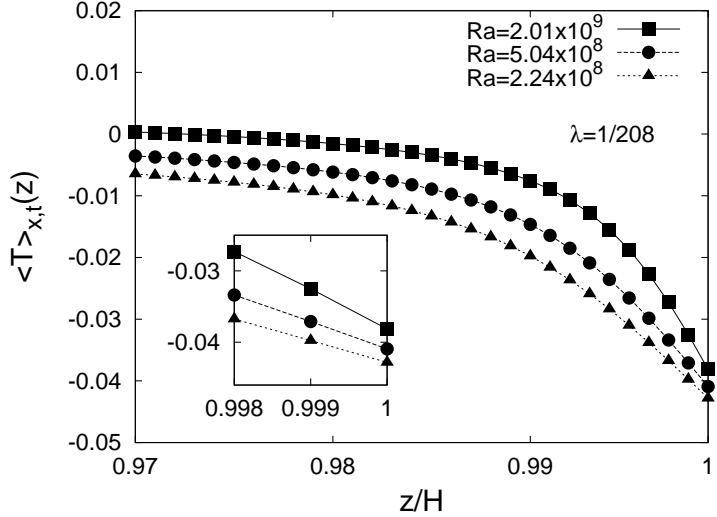}
\caption{Vertical temperature profile (top boundary) for fixed $\lambda=1/208$ and $\xi=0.5$, at changing $Ra$: $Ra=2.24 \times 10^8$ ($\blacktriangle$), $Ra=5.04 \times 10^8$ ($\bullet$), $Ra=2.01 \times 10^9$ ($\blacksquare$). Inset: enlargement of the region close to the top boundary. }
\label{fig:blu}
\end{center}
\end{figure}

In figure \ref{fig:blu} we show the effects of the thermal corrugation on the averaged temperature profile $\langle T \rangle_{x,t}$ at the upper wall for three different Rayleigh numbers. Notice that at increasing $Ra$, the thermal boundary layer becomes thinner (as expected) but also the 'effective' mean temperature at the wall increases, because the insulating region introduces a perturbation that is larger and larger with respect to the width of the thermal boundary layer. Our numerical resolution is such that even at the highest Rayleigh numbers investigated we have enough grid points in the boundary layer to observe smooth profiles. In table \ref{tab:1} we summarized all  details of the numerical set up.  



\begin{center}
\begin{figure}
\includegraphics[scale=0.56,angle=0]{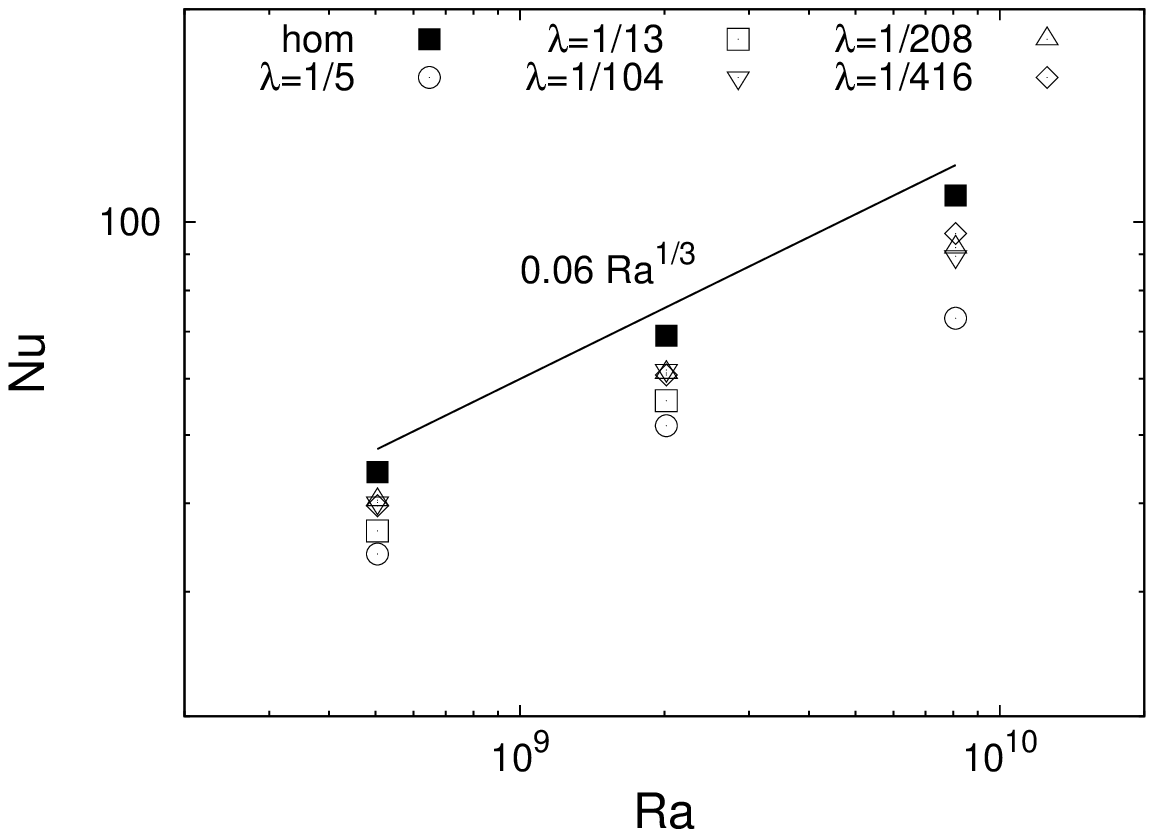}
\includegraphics[scale=0.56,angle=0]{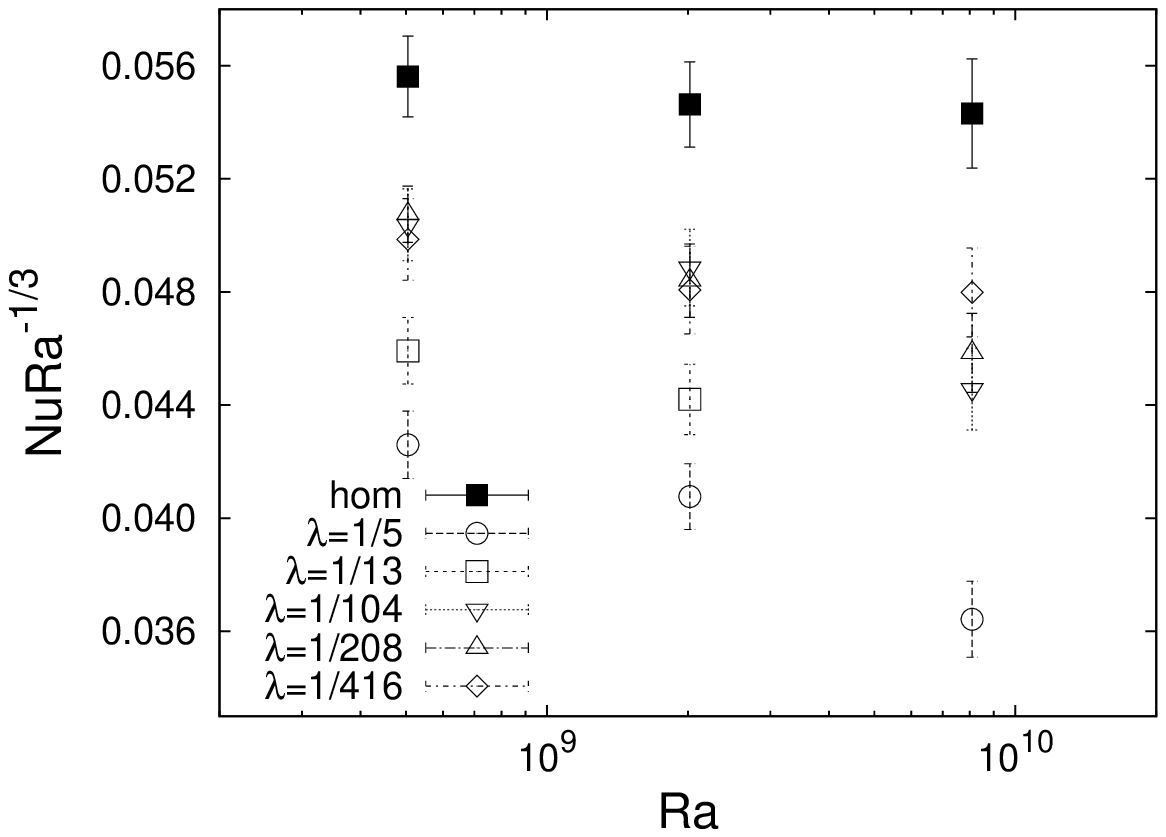}
\caption{Left: $Nu$ as a function of $Ra$ at various $\lambda$ for fixed $L_{x}/H=2$ and for fixed $\xi=0.5$: $\lambda=1/5$ ($\circ$), $\lambda=1/13$ ($\boxdot$), $\lambda=1/104$ ($\triangledown$), $\lambda=1/208$ ($\vartriangle$), $\lambda=1/416$ ($\lozenge$). The fully homogeneous case is also shown for comparison ($\blacksquare$). Right: same data of left panel normalized with the $\sim Ra^{1/3}$ law. The dependency on $\lambda$ is more pronounced at increasing $Ra$, because the thermal boundary layer thickness, $\lambda_T$, gets smaller than the period of the boundary pattern, $L$ (see figure \ref{fig:lambda}). \label{fig:nusselt}}
\end{figure}
\end{center}

\begin{figure}
\begin{center}
\includegraphics[scale=0.8,angle=0]{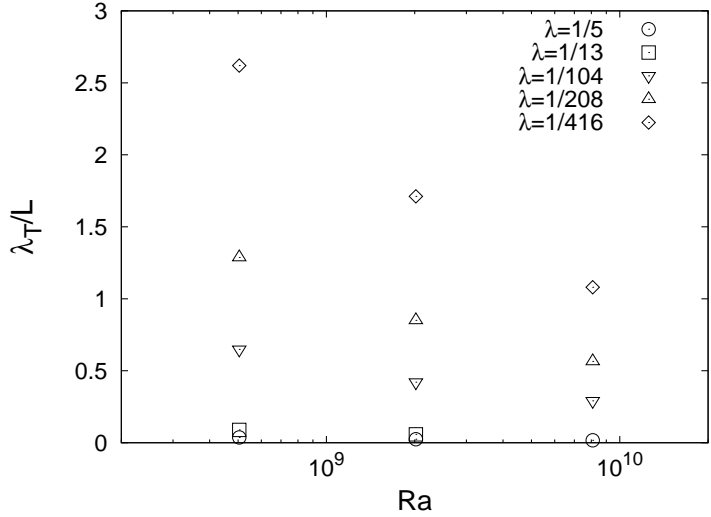}
\caption{Ratio between the thermal boundary layer and the period of the boundary pattern, $\frac{\lambda_T}{L}$, as a function of $Ra$. The thermal boundary layer has been computed from the Nusselt number, i.e.  $\lambda_T=H/ (2 Nu)$. Different values of $\lambda$ are considered: $\lambda=1/5$ ($\circ$), $\lambda=1/13$ ($\boxdot$), $\lambda=1/104$ ($\triangledown$), $\lambda=1/208$ ($\vartriangle$), $\lambda=1/416$ ($\lozenge$).}
\label{fig:lambda}
\end{center}
\end{figure}

\begin{center}
\begin{figure}
\hspace{-9mm}
\vspace{-17mm}
\includegraphics[scale=0.45,angle=0]{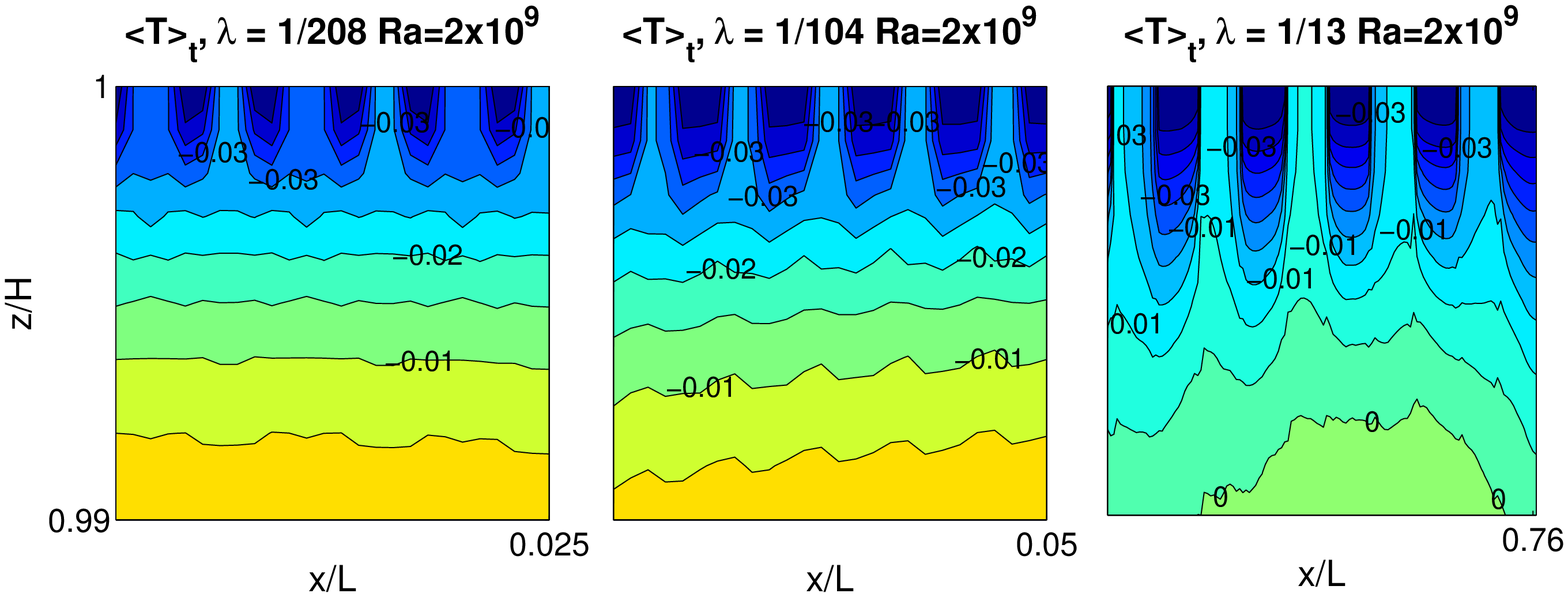}
\vspace{-8mm}
\hspace{-9mm}
\includegraphics[scale=0.45,angle=0]{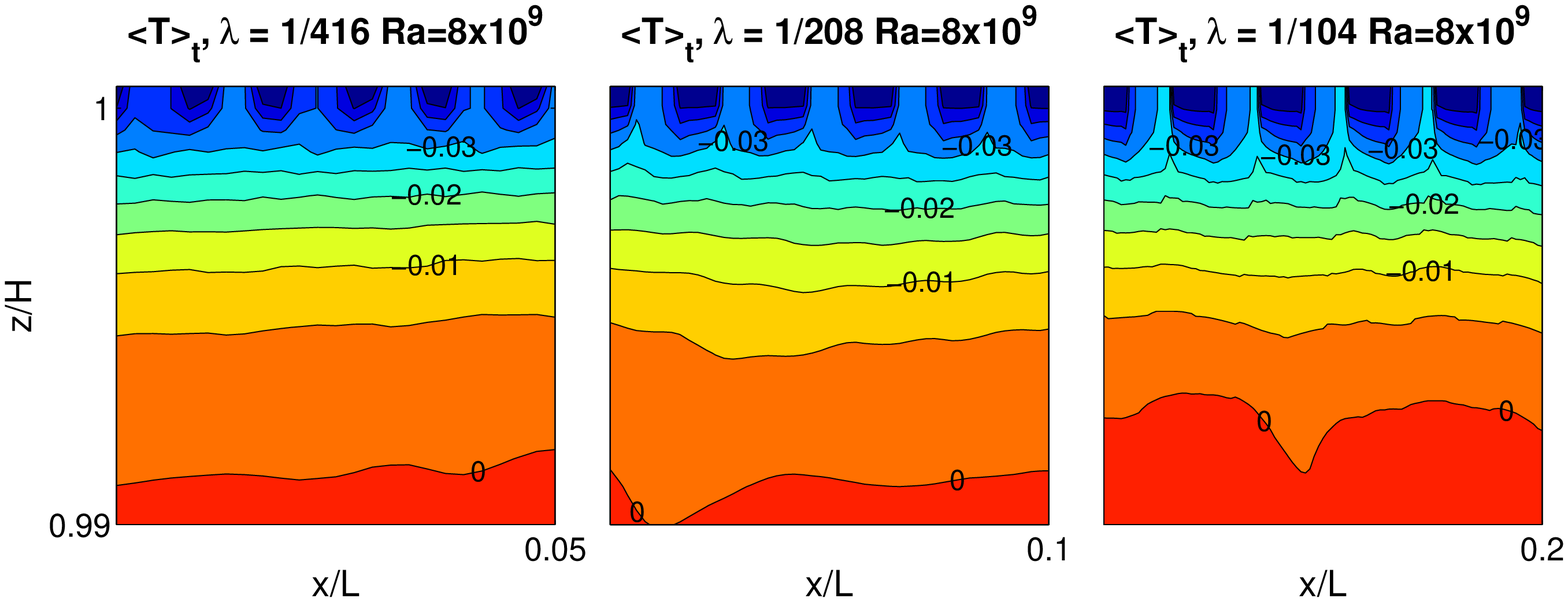}
\caption{(Color online) Temperature profile averaged in time, $\langle ... \rangle_t$, close to a block of insulating/conducting regions in the top boundary. Top row from right to left: $Ra=2 \times 10^9$, $\lambda= 1/13,1/104,1/208$. Bottom row from right to left: $Ra=8 \times 10^9$, $\lambda= 1/104, 1/208,1/416$. Notice that in the top row the two profiles at $\lambda=1/104,1/208$ have a thermal corrugation length of the same order of (or smaller than) the typical thermal boundary layer width, $\lambda_T$, and therefore their normalized Nusselt is not changing any more (see right panel of figure \ref{fig:nusselt}). For the bottom row, only the case at $\lambda=1/416$ has a corrugation of the order of $\lambda_T/2$, and the Nusselt number of the three cases still shows some {\it small} dependency on $\lambda$.} 
\label{fig:plot2d}
\end{figure}
\end{center}

Concerning the heat exchange properties, in the left panel of figure \ref{fig:nusselt} we show the Nusselt number as a function of Rayleigh number;  in the right panel, instead, we normalize the Nusselt number with the $\sim Ra^{1/3}$ empirical rule observed for homogeneous RB systems (see, e.g. \cite{Ahlers,Chilla} for detailed discussions about possible corrections to the 
dimensional $1/3$ law). From these figures one can observe the two main effects already discussed before. Looking on a global scale (left panel), we do not observe any strong effect of the boundary non-homogeneities on the heat exchange, at least as far as the scaling properties of $Nu$ vs. $Ra$ are concerned. At a closer look (right panel), after compensation with $Ra^{1/3}$, some {\it small effects can be  indeed detected}.  First, let us fix $Ra$ and look at what happens at increasing $\lambda$ by keeping constant $\xi=0.5$. For example, for $Ra \sim 5\times 10^8$, we observe a systematic increase of the heat flux by decreasing $\lambda$ up to a critical value of the boundary corrugation where nothing changes anymore by keeping reducing it. This is clearly in agreement with the statement that thermal corrugations in the boundary might affect the bulk physics only when their typical length is larger than -or of the order of-  the thermal boundary layer. Similarly, moving at higher Rayleigh number, say $Ra \sim 10^{10}$, we still observe a discrepancy between the heat fluxes even for those values of $\lambda$ that already had saturated at a lower $Ra$. This is due to the fact that increasing $Ra$ decreases $\lambda_T$ and that therefore those patterns that satisfy $L \ll \lambda_T$ at a low $Ra$, do not satisfy it anymore for higher $Ra$. It is difficult to quantify this argument and the transition cannot be sharp. Mainly because the very definition of $\lambda_T$ depends also  on the control parameters $\lambda$ and $\xi$. In figure \ref{fig:lambda} we show the behavior of $\lambda_T/L $  for the different data sets of figure \ref{fig:nusselt} in order to give a qualitative support to the previous statement. Indeed, we see that when $\lambda_T/L \sim 0.8$  or larger, the bulk heat transfer seems to become independent of the corrugation details. Finally, in figure \ref{fig:plot2d} we show a typical measurement of the mean temperature profile close to the upper boundary for different $\lambda$ at a given $Ra$, where we can see indeed that whenever $Nu$ does not depend anymore on $\lambda$, we also observe that $\lambda_T>L$.
 

\begin{table*}
\begin{tabular}{c|cccccccccc}
  \hline 
  \hline 
  $Ra$ & $\lambda$ & $L_x$ & $H$ & $\nu$ & $\kappa$ & Pr & $\Delta T$ & $g$ & $\tau_{RB}$ & $t_{tot}/{\tau_{RB}}$ \\ 
  \hline
  \hline  
  5 $\times 10^8$ & hom & 2080 & 1040 & $3.3 \times 10^{-3}$ & $3.3 \times 10^{-3}$ & 1 & 0.1  & $5 \times 10^{-5}$ & $2.5 \times 10^{5}$ & 26 \\
  5 $\times 10^8$ & 1/5 & 2080 & 1040 & $3.3 \times 10^{-3}$ & $3.3 \times 10^{-3}$ & 1 & 0.1  & $5 \times 10^{-5}$ & $2.5 \times 10^5$& 40 \\
  5 $\times 10^8$ & 1/13 & 2080 & 1040 & $3.3 \times 10^{-3}$ & $3.3 \times 10^{-3}$ & 1 & 0.1  & $5 \times 10^{-5}$ & $2.5 \times 10^5$& 40 \\
  5 $\times 10^8$ & 1/208 & 2080 & 1040 & $3.3 \times 10^{-3}$ & $3.3 \times 10^{-3}$ & 1 & 0.1  & $5 \times 10^{-5}$ & $2.5 \times 10^5$ & 40 \\
  5 $\times 10^8$ & 1/416 & 4160 & 2080 & $1.6 \times 10^{-3}$ & $1.6 \times 10^{-3}$ & 1 & 0.1  & $1.5 \times 10^{-6}$ & $2.8 \times 10^6$ & 45 \\

\hline

  2 $\times 10^9$ & hom & 2080 & 1040 & $1.6 \times 10^{-3}$ & $1.6 \times 10^{-3}$ & 1 & 0.1  & $3.5 \times 10^{-5}$ & $3.5 \times 10^{5}$ & 24 \\
  2 $\times 10^9$ & 1/5 & 2080 & 1040 & $1.6 \times 10^{-3}$ & $1.6 \times 10^{-3}$ & 1 & 0.1  & $3.5 \times 10^{-5}$ & $3.5 \times 10^5$& 42 \\
  2 $\times 10^9$ & 1/13 & 2080 & 1040 & $1.6 \times 10^{-3}$ & $1.6 \times 10^{-3}$ & 1 & 0.1  & $3.5 \times 10^{-5}$ & $3.5 \times 10^5$& 42 \\
  2 $\times 10^9$ & 1/208 & 2080 & 1040 & $1.6 \times 10^{-3}$ & $1.6 \times 10^{-3}$ & 1 & 0.1  & $3.5 \times 10^{-5}$ & $3.5 \times 10^5$ & 42 \\
  2 $\times 10^9$ & 1/416 & 4160 & 2080 & $1 \times 10^{-3}$ & $1 \times 10^{-3}$ & 1 & 0.1  & $1.5 \times 10^{-6}$ & $3.7 \times 10^6$ & 45 \\

\hline

  8 $\times 10^9$ & hom & 4160 & 2080 & $1.6 \times 10^{-3}$ & $1.6 \times 10^{-3}$ & 1 & 0.1  & $2.5 \times 10^{-6}$ & $7 \times 10^{5}$ & 20 \\
  8 $\times 10^9$ & 1/5 & 4160 & 2080 & $1.6 \times 10^{-3}$ & $1.6 \times 10^{-3}$ & 1 & 0.1  & $2.5 \times 10^{-6}$ & $7 \times 10^5$& 47 \\
  8 $\times 10^9$ & 1/208 & 4160 & 2080 & $1.6 \times 10^{-3}$ & $1.6 \times 10^{-3}$ & 1 & 0.1  & $2.5 \times 10^{-6}$ & $7 \times 10^5$ & 47 \\
  8 $\times 10^9$ & 1/416 & 4160 & 2080 & $1.6 \times 10^{-3}$ & $1.6 \times 10^{-3}$ & 1 & 0.1  & $2.5 \times 10^{-6}$ & $7 \times 10^5$ & 47 \\

  \hline 
  \hline 

\end{tabular}
\label{tab:1}
\caption{Numerical parameters for all simulations performed with the LBM algorithm. Different columns refers to: Rayleigh number, $Ra= g \Delta T H^3/ (\nu \kappa)$; pattern length in units of the cell height, $\lambda = L/H$; the horizontal periodic length, $L_x$;  viscosity and thermal diffusivity, $\nu = c_s^2(\tau_f-0.5)$, $\kappa= c_s^2(\tau_g-0.5)$, where $c_s^2=1/3$ is the sound speed and $\tau_f$ and $\tau_g$ are the relaxation times in the Boltzmann equation (see text); Prandtl number; difference between the bottom plate and top plate imposed temperatures $\Delta T = (T_{down}-T_{up})$; gravity, $g$; large scale eddy turn over time, $\tau_{RB} = \sqrt{\frac{L_x}{{\alpha}g{\Delta}T}}$; total integration time, $t_{tot}/\tau_{RB}$.}
\end{table*}


\section{Conclusions} \label{s:concl}

 Natural convection with non-homogeneous horizontal boundary conditions has been investigated in presence of alternating strips of conducting and insulating boundary regions. The  simplest, one-dimensional, geometrical patterning has been investigated at changing the (i) Rayleigh number, (ii)  insulating/conducting surface ratio  and (iii) patterning periodicity. For moderate Rayleigh numbers we presented both analytical and numerical evidences that the transition to bulk convection is delayed with respect to the homogeneous case, happening  at a renormalized Rayleigh number which depends on the patterning properties. At high Rayleigh numbers we used numerical simulations based on a lattice Boltzmann method for a two dimensional horizontally periodic box to show that the control parameter can be identified in  the ratio between the thermal  boundary layer width, $\lambda_T$, and the characteristic period $L$ of the boundary pattern. For $L \gg \lambda_T$, the insulating regions give an effective corrugation in the boundary and decreases the normalized heat flux. The net effect, comparing two patterning with periodicity differing of almost two order of magnitude (i.e. $\lambda = 1/5, 1/416$), can lead to an enhancement in the normalized heat flux by a factor $40-50\%$.  The above findings show that for a given surface ratio of insulating/conducting regions it is more efficient to use tiles that have a characteristic size smaller than -or equal to-  the boundary layer thickness.  The situation is  different to the case of rough geometrical elements on the surface, which lead to an increase of heat transport when the characteristic size of the roughness elements is larger than the boundary layer and penetrate it (see \cite{castaing,wagner} for experimental and theoretical studies addressing the case of multi-scale roughness or regular patterning, respectively). Clearly, for any fixed pattern distribution, there will always exist a Rayleigh number high enough such that the non-homogeneous structure will emerge. Further numerical investigations at changing the insulating/conducting distribution and pattern, and involving also 3D geometries would be very welcome to understand the robustness of such finding in the general case. Even more interesting could be the case where the patterning shows multi-scale non-homogeneous properties as for the case of ice covering in the ocean. Another further direction of investigation would be to add inhomogeneities also at the bottom plate and gauge the formation of stable large scale flows triggered by a preferential patterning.  Finally, following the same approach, it could be interesting to study the importance of adding also time modulation in the boundary conditions to detect possible  synchronization effects in the detachment of plumes and highlighting an optimal forcing protocol to enhance heat transfer between the two plates following the  works of \cite{jin,lohsepre}. \\
The authors kindly acknowledge funding from the European Research Council under the EU Seventh Framework Programme (FP7/2007-2013) / ERC Grant Agreement no[279004]. We acknowledge computational support from CINECA (IT).

\section{Appendix A}
In this section we detail the calculations to solve the DS in the limit $L\ll H$. The general expression of such a DS is
\be\label{DUAL1}
\begin{split}
\begin{cases}
\frac{a_0}{2} +\sum_{j=1}^{\infty} a_j  \cos \left( j \tilde{x} \right)=0 & c<\tilde{x}<\pi \\
\sum_{j=1}^{\infty} j a_j   \cos \left( j \tilde{x} \right)=f(\tilde{x}) & 0<\tilde{x}<c,
\end{cases}
\end{split}
\ee
with $f(\tilde{x}) =1$. The solution of eqs. (\ref{DUAL1}) can be found on page 161 of the book by \citet{sneddon66} in terms of a function $h_1(t)$:
\be\label{a0}
\begin{cases}
a_{0}=\frac{2}{\pi}\left[ \frac{\pi}{\sqrt{2}}\displaystyle\int_{0}^{c}h_{1}(t) dt \right]\\
a_{j}=\frac{2}{\pi}\left[ \frac{\pi}{2\sqrt{2}}\displaystyle\int_{0}^{c}h_{1}(t)[P_{j}(\mbox{cos} (t) )+P_{j-1}(\mbox{cos} (t))] dt \right] \hspace{.2in}  j=1,2,...
\end{cases}
\ee
where $P_{j}$ are the Legendre polynomials. The function $h_{1}(t)$ is such that
\be\label{h}
h_{1}(t)=\frac{2}{\pi}\frac{d}{d t} \displaystyle\int_{0}^{t}\frac{\mbox{sin} (\frac{x}{2}) ~ dx}{\sqrt{\mbox{cos}(x)-\mbox{cos}(t)  }}
\displaystyle \left( \int_{0}^{x}f(u)du \right)
\ee
and, in our case where $f(\tilde{x})=1$, we get 
\be\label{h0}
h_{1}(t)=\frac{2}{\pi} \frac{d}{d t} \displaystyle\int_{0}^{t}\frac{x\,\mbox{sin} (\frac{x}{2}) ~ dx}{\sqrt{\mbox{cos}(x)-\mbox{cos}(t)  }} \, .
\ee
The computation of $a_{0}$ hinges  on the knowledge of the function $h_{1}(t)$ which is the derivative of the integral
\be\label{I}
I(t)=\frac{2}{\pi} \displaystyle\int_{0}^{t}\frac{x ~ \mbox{sin} (\frac{x}{2}) ~ dx}{\sqrt{\mbox{cos}(x)-\mbox{cos}(t)  }}.
\ee 
The integral in (\ref{I}) can be evaluated with some manipulations and the use  of formula 3.842 of  the book by \citet{GR00}, leading to:
\be
I(t)= \frac{4 }{\sqrt{2}} \mbox{log}\left( \frac{1}{\mbox{cos}(\frac{t}{2}) }\right)
\ee
from which we get $h_{1}(t)= \sqrt{2} \mbox{tan} \left( \frac{t}{2} \right)$ and, consequently, the exact expression for $a_0$:
\be
a_0=4 \mbox{log}\left( \frac{1}{\mbox{cos}(\frac{c}{2})} \right)
\ee
where we have used $c=\pi \xi$ with $\xi$ the insulating fraction. The other coefficients can also be found using formula (\ref{a0}). Although not necessary for the scope of the paper, it may be of interest to also give the expression for the first coefficients beside $a_0$:
\be\label{a1}
\begin{cases}
a_1=2 \sin^2 \left( \frac{c}{2} \right)\\
a_2=-1-3 \cos^4 \left( \frac{c}{2} \right)+4\cos^2 \left( \frac{c}{2} \right)\\
a_3=\frac{2}{3}-\frac{20}{3} \cos^6 \left( \frac{c}{2} \right)+12\cos^4 \left( \frac{c}{2} \right)-6\cos^2 \left( \frac{c}{2} \right).\\
\end{cases}
\ee
We also notice that using the results of \citet{sneddon66} we can analytically solve for the profile of the temperature in the insulating region at the top wall, $0<x<c$:
\be\label{VEL0}
\frac{1}{2}a_{0}+\sum_{j=1}^{\infty} a_{j} \mbox{cos}(j \tilde{x})= \mbox{cos}{\left(\frac{\tilde{x}}{2}\right)}\displaystyle\int_{\tilde{x}}^{c}\frac{h_{1}(t) dt}{\sqrt{\mbox{cos}(x)-\mbox{cos}(t)}}.
\ee
The integral in expression (\ref{VEL0}) can be done exactly to get the temperature at the upper insulating region:
\be\label{V0}
\frac{1}{2}a_{0}+\sum_{j=1}^{\infty} a_{j} \mbox{cos}(j \tilde{x})=2  \mbox{arccosh}\left(\frac{\mbox{cos}(\frac{\tilde{x}}{2})}{\mbox{cos}(\frac{\xi \pi}{2})} \right) \hspace{.2in} 0<\tilde{x}<c
\ee
that is the expression reported in (\ref{profH}) and used to benchmark our numerically assisted solution for the DS reported in section \ref{sec:dualSeries}.

\section{Appendix B}

In this section we give the explicit expression for the coefficients characterizing the DS equation in the intermediate case, $1 \approx \frac{L}{H} \gg e^{-4 \pi H/L}$. The DS reported in eq. (\ref{INTERMEDIATE}) can be rewritten as
\be
\begin{split}
\begin{cases}
\frac{a_0}{2} +\sum_{j=1}^{\infty} a_j  \cos \left( j \tilde{x} \right)=0 & c< \tilde{x}<\pi \\
\frac{a_0 \lambda}{4 \pi} + \sum_{j=1}^{\infty} j a_j   \cos \left( j \tilde{x} \right)=1 & 0< \tilde{x}<c
\end{cases}
\end{split}
\ee
with $\lambda=\frac{L}{H}$. We use the general solution given by eqs. (\ref{a0}) and (\ref{h}), with the function $h_{1}(t)$ given by
\be\label{hi}
h_{1}(t)=\frac{2}{\pi}\frac{d}{d t} \displaystyle\int_{0}^{t}\frac{\mbox{sin} (\frac{x}{2}) ~ dx}{\sqrt{\mbox{cos}(x)-\mbox{cos}(t)  }}
\displaystyle \left( \int_{0}^{x}f(u) du - \frac{\lambda}{4 \pi} a_0 x \right).
\ee
The integrals are the same of  the previous Appendix. This leads to the following expression for $h_1(t)$: 
\be
h_{1}(t)= \sqrt{2}\left(1-\frac{\lambda}{4 \pi} a_0 \right) \mbox{tan} \left( \frac{t}{2} \right) 
\ee
and an equation for $a_0$:
\be
a_0=4 \left(1-\frac{\lambda}{4 \pi} a_0 \right) \mbox{log}\left( \frac{1}{\mbox{cos}(\frac{c}{2})} \right).
\ee
Therefore, we determine $h_1$ and $a_0$ as follows
\be
h_{1}(t)=\sqrt{2}\left(1-\frac{L}{4\pi H} \frac{4  \mbox{log}\left( \frac{1}{\mbox{cos}(\frac{c}{2})} \right)}{1+\frac{L}{ \pi H} \mbox{log}\left( \frac{1}{\mbox{cos}(\frac{c}{2})} \right)} \right) \mbox{tan} \left( \frac{t}{2} \right) 
\ee
\be
a_0=\frac{4  \mbox{log}\left( \frac{1}{\mbox{cos}(\frac{c}{2})} \right)}{1+\frac{\lambda}{\pi} \mbox{log}\left( \frac{1}{\mbox{cos}(\frac{c}{2})} \right)}=\frac{4  \mbox{log}\left( \frac{1}{\mbox{cos}(\frac{c}{2})} \right)}{1+\frac{L}{\pi H} \mbox{log}\left( \frac{1}{\mbox{cos}(\frac{c}{2})} \right)}.
\ee
Besides $a_0$, the first three coefficients are found to be:
\be\label{a1i}
\begin{cases}
a_1=\left(1-B(c)\right)\left[ 2 \sin^2 \left( \frac{c}{2} \right) \right]\\
a_2=\left(1-B(c)\right) \left[ 1-3 \cos^4 \left( \frac{c}{2} \right)+4\cos^2 \left( \frac{c}{2} \right) \right]\\
a_3=\left(1-B(c)\right) \left[ \frac{2}{3}-\frac{20}{3} \cos^6 \left( \frac{c}{2} \right)+12\cos^4 \left( \frac{c}{2} \right)-6\cos^2 \left( \frac{c}{2} \right) \right]
\end{cases}
\ee         
where, for simplicity, we have defined the function 
$$
B(c) =  \frac{L}{4\pi H} \frac{4  \mbox{log}( \frac{1}{\mbox{cos}(\frac{c}{2})} )}{1+\frac{L}{ \pi H} \mbox{log}( \frac{1}{\mbox{cos}(\frac{c}{2})} )}.
$$

\section{Appendix C}
\begin{figure*}
\begin{center}
\includegraphics[scale=.5]{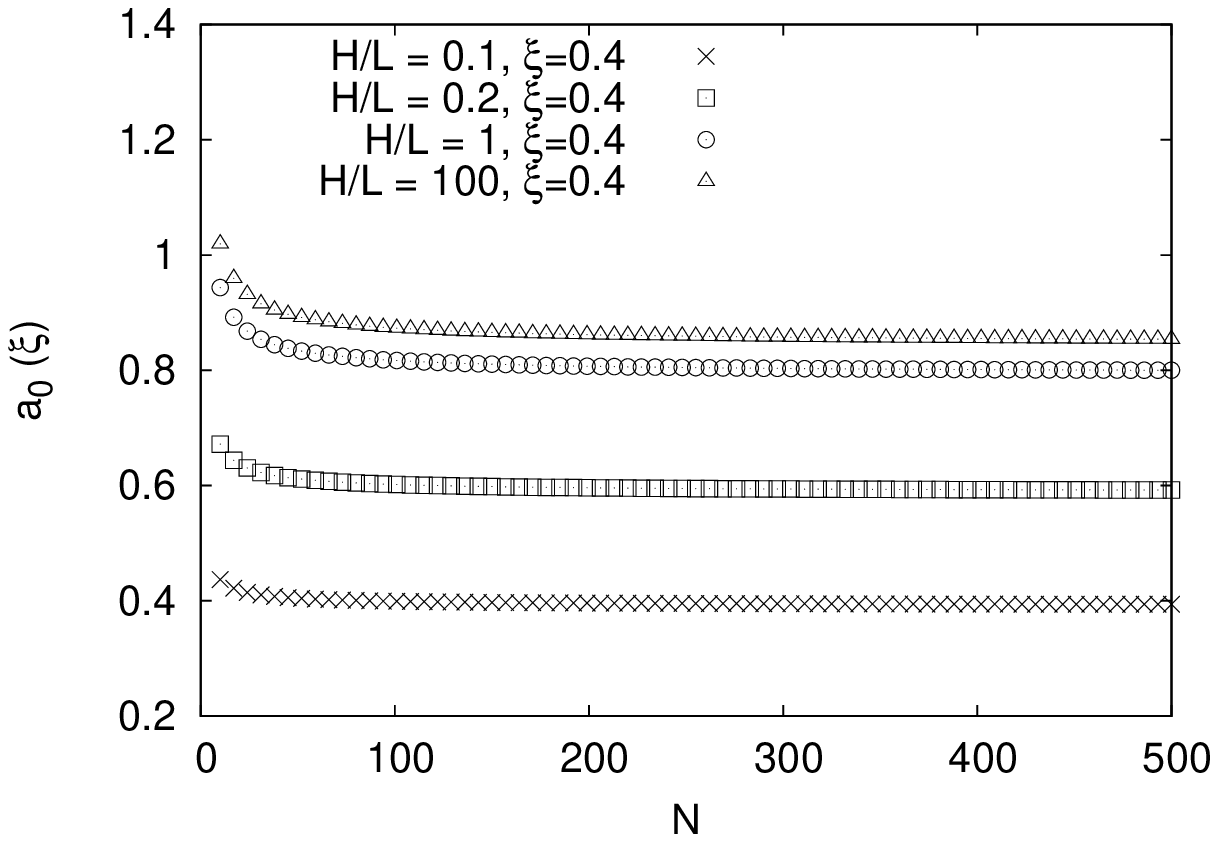}
\includegraphics[scale=.5]{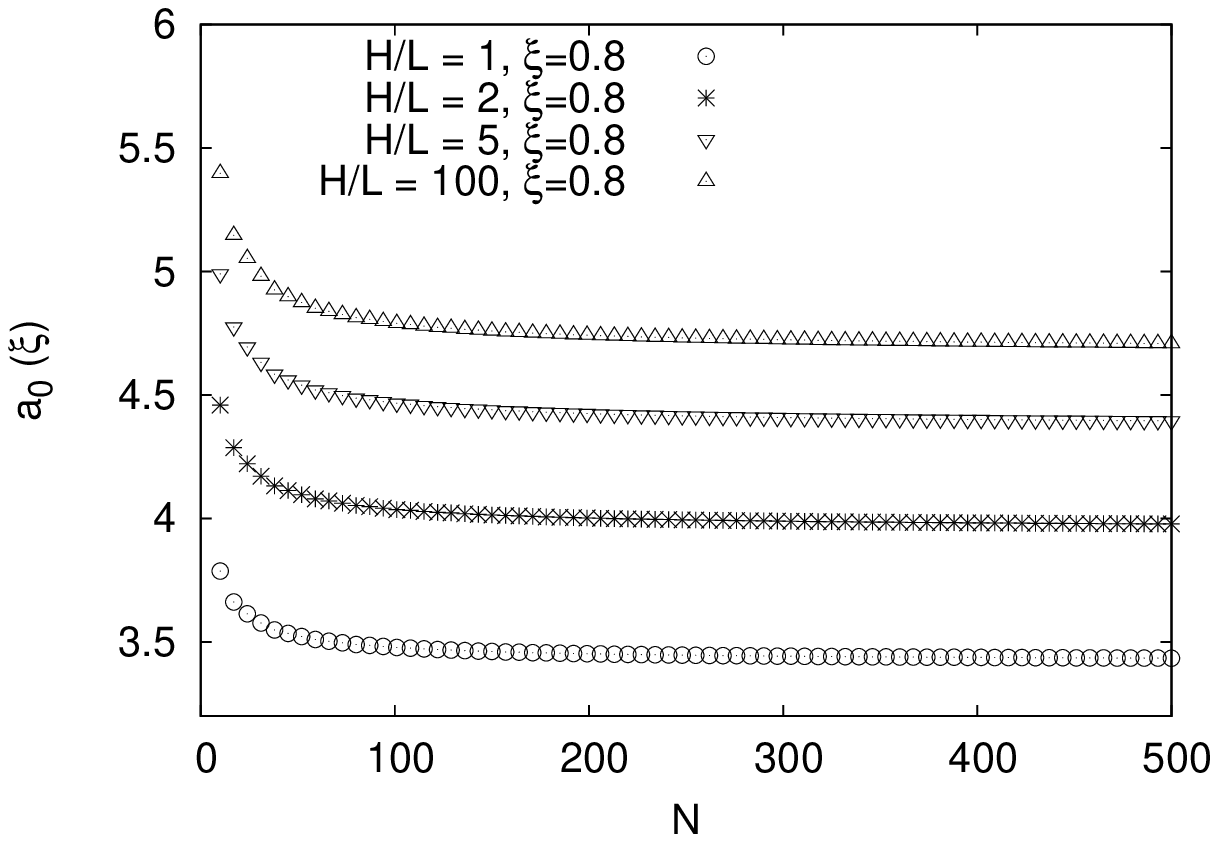}
\vspace{-3mm}
\caption{Convergence rate of the DS at changing the number of terms $N$ in the expansion.}
\label{fig:trunc}
\end{center}
\end{figure*}

In this appendix we report the details of the linear system used to solve the general DS problem of section \ref{sec:dualSeries}. The starting point is the equation
\be\label{GENERALDS}
\sum_{j=0}a_{j} F_{j}(\tilde{x}) \cos(j\tilde{x})=G(\tilde{x})
\ee
with $G(\tilde{x})=\theta(\tilde{x})\theta(c-\tilde{x})$ and
\be
F_{j}(\tilde{x})=\left \{  \begin{array}{l l} \frac{1}{2}\theta(\tilde{x}-c)\theta(\pi-\tilde{x}) + \left( \frac{L}{4 \pi H} \right)  \theta(\tilde{x}) \theta(c-\tilde{x})  & j=0  \\  \left(1- e^{-\frac{4 \pi j H}{L}} \right) \theta(\tilde{x}-c)\theta(\pi-\tilde{x})+ j   \left(1+ e^{-\frac{4 \pi j H}{L}} \right)\theta(\tilde{x})\theta(c-\tilde{x})             & j>0. \\
\end{array} \right.
\ee
Taking the inner product with $\cos (i \tilde{x})$ we  reduce the problem  to the  linear system: 
\be\label{LINSYS}
A_{i,j} a_{j}=y_{i}.
\ee
The function $F_{j}(\tilde{x})$ in (\ref{GENERALDS}) is piece-wise constant and, upon multiplying by $\cos (i \tilde{x})$ and integrating in the interval $0\le \tilde{x}\le \pi$, we get:
\be
\begin{split}
\displaystyle\int_{c}^{\pi} \frac{a_{0}}{2}  \mbox{cos} (i \tilde{x}) d\tilde{x}  + & \sum_{j=1}^{\infty} a_{j}\left(1- e^{-\frac{4 \pi j H}{L}} \right) \displaystyle\int_{c}^{\pi} \mbox{cos} (j \tilde{x}) \mbox{cos} (i \tilde{x}) d \tilde{x} + \displaystyle\int_{0}^{c}  a_0 \frac{L}{4 \pi H} \mbox{cos} (i \tilde{x}) d \tilde{x} \\ 
+ & \sum_{j=1}^{\infty} j a_{j}\left(1+ e^{-\frac{4 \pi j H}{L}} \right) \displaystyle\int_{0}^{c}  \mbox{cos} (i \tilde{x}) \mbox{cos} (j \tilde{x}) d \tilde{x}-  \displaystyle\int_{0}^{c} \mbox{cos} (i \tilde{x})  d\tilde{x}=0. 
\end{split}
\ee
Using the following integral:
\be
\displaystyle\int_{0}^{c} \mbox{cos} (i \tilde{x})  \mbox{cos} (j \tilde{x}) d \tilde{x} = 
\left \{  \begin{array}{l l} -\displaystyle\int_{c}^{\pi} \mbox{cos} (i \tilde{x}) \mbox{cos} (j \tilde{x}) d \tilde{x}  & i \neq j \\
-\displaystyle\int_{c}^{\pi} \mbox{cos} (i \tilde{x}) \mbox{cos} (j \tilde{x}) d\tilde{x} +\frac{\pi}{2}& i=j \\
\end{array} \right.
\ee
we obtain: 
\be
\begin{split}
& \displaystyle\int_{c}^{\pi} \frac{a_{0}}{2}  \mbox{cos} (i \tilde{x}) d \tilde{x}  + \sum_{j=1}^{\infty} a_{j} \left(1- e^{-\frac{4 \pi j H}{L}} \right) \left(- \displaystyle\int_{0}^{c} \mbox{cos} (i \tilde{x}) \mbox{cos} (j \tilde{x}) d \tilde{x} +  \frac{\pi}{2}\delta_{ij} \right)  \\
+&\displaystyle\int_{0}^{c}  a_0 \frac{L}{4 \pi H} \mbox{cos} (i \tilde{x}) d \tilde{x}+ \sum_{j=1}^{\infty} j a_{j}\left(1+ e^{-\frac{4 \pi j H}{L}} \right)  \displaystyle\int_{0}^{c}  \mbox{cos} (i \tilde{x}) \mbox{cos} (j \tilde{x}) d  \tilde{x}=  \displaystyle\int_{0}^{c} \mbox{cos} (i \tilde{x})  d \tilde{x}.
\end{split}
\ee
The term $y_i$ in (\ref{LINSYS}) is therefore given by
\be
y_{i}= \displaystyle\int_{0}^{c} \mbox{cos} (i \tilde{x}) d \tilde{x}=\left \{  \begin{array}{l l} c  & i = 0  \\
                                                     \frac{1}{i} \sin(ic) & i \neq 0 . \\
\end{array} \right. 
\ee
To write out the matrix elements $A_{i,j}$ we need some algebra. An indefinite integral of interest is
$$
\int \cos (i \tilde{x}) \cos (j \tilde{x}) d \tilde{x} = \frac{1}{2} \frac{\sin ((i-j) \tilde{x})}{i-j}+\frac{1}{2} \frac{\sin ((i+j) \tilde{x})}{i+j}+\mbox{const.}
$$
so that we estimate
\be
\int_0^c \cos (i \tilde{x}) \cos (j \tilde{x}) d \tilde{x} =\left \{  \begin{array}{l l} \frac{1}{2} \frac{\sin ((i-j)c)}{i-j}+\frac{1}{2} \frac{\sin ((i+j)c)}{i+j}  & i \neq j  \\
                                                \frac{c}{2}+    \frac{1}{2} \frac{\sin ((i+j)c)}{i+j}  & i=j. \\
\end{array} \right.
\ee
Finally, we need 
\be
\displaystyle\int_{c}^{\pi} \mbox{cos} (i \tilde{x}) d \tilde{x}=\left \{  \begin{array}{l l} \pi-c  & i = 0  \\
                                                     -\frac{1}{i} \sin(ic) & i \neq 0. \\
\end{array} \right.
\ee
To summarize, we obtain 
\be
\begin{split}
A_{i,j} = & \delta_{i0}\delta_{j0} \frac{1}{2}\left(\frac{L}{2 \pi H}c +\pi -c  \right)+ (1-\delta_{i0})\delta_{j0}\frac{1}{2}\left(\frac{L}{2 \pi H} -1  \right)\frac{1}{i} \sin(ic) \nonumber \\
+& \delta_{ij}(1-\delta_{j0}) \left[ \left( j\left(1+ e^{-\frac{4 \pi j H}{L}} \right)-\left(1- e^{-\frac{4 \pi j H}{L}} \right) \right) \left( \frac{c}{2}+\frac{1}{2} \frac{\sin ((i+j)c)}{i+j} \right)+\left(1- e^{-\frac{4 \pi i H}{L}} \right) \frac{\pi}{2} \right] \nonumber\\
+& (1-\delta_{ij})(1-\delta_{j0})\left[ \left( j\left(1+ e^{-\frac{4 \pi j H}{L}} \right)-\left(1- e^{-\frac{4 \pi j H}{L}} \right) \right) \left( \frac{1}{2} \frac{\sin ((i-j)c)}{i-j}+\frac{1}{2} \frac{\sin ((i+j)c)}{i+j} \right) \right].
\end{split}
\ee
In order to give a quantitative idea of the rate of convergence of the DS as a function of the truncation order $N$, we report in figure \ref{fig:trunc} results concerning $a_0$ for two different boundary conditions and different cell aspect ratios. 

\bibliographystyle{jfm}

\end{document}